\RequirePackage{fix-cm}
\RequirePackage{rotating}

\documentclass{svjour3}                     

\smartqed
\usepackage[utf8]{inputenc}
\usepackage{breakcites}
\usepackage{graphicx}
\usepackage{tabularx}
\usepackage{diagbox}
\usepackage{booktabs}
\usepackage{natbib}
\usepackage{makecell}
\usepackage{multirow}
\usepackage{multicol}
\usepackage{listings}
\setcitestyle{aysep={}}
\usepackage{xcolor, soul}
\usepackage{latexsym}
\usepackage{makeidx}
\usepackage{hyperref}
\usepackage{empheq}
\usepackage{caption}
\usepackage{array}
\usepackage{setspace}
\usepackage{spverbatim}
\usepackage{import}
\usepackage{subcaption}
\usepackage{amsmath}
\usepackage{tikz}
\usetikzlibrary{calc,trees,positioning,arrows,chains,shapes.geometric,decorations.pathreplacing,decorations.pathmorphing,shapes,matrix,shapes.symbols,shadows}

\usepackage{mdframed}
\usepackage{boxedminipage}

\newcolumntype{R}{>{\raggedleft\arraybackslash}X}
\newcolumntype{L}{>{\raggedright\arraybackslash}X}

\definecolor{mygray}{HTML}{CED4E0}
\sethlcolor{mygray}
\definecolor{FrameColor}{rgb}{0.93,0.93,0.93}
\definecolor{ao}{rgb}{0.0, 0.5, 0.0}
\definecolor{yellow}{rgb}{0.89, 0.82, 0.04}
\lstdefinestyle{tblstyle}{
    language=C,
    basicstyle={\ttfamily},
    breaklines=true,
    breakatwhitespace=true,
    postbreak=\mbox{\textcolor{gray}{$\hookrightarrow$}\space},
    frame=none,
    keepspaces=true,
    aboveskip=-8pt,
    belowskip=-10pt,
    mathescape=true
}
\tikzset{
    >=stealth',
    process/.style={draw=black,very thick,circle,fill=white,text centered,},
    punkt/.style={
           rectangle,
           rounded corners,
           draw=black, very thick,
           text width=6.5em,
           minimum height=2em,
           text centered},
    pil/.style={
           ->,
           thick,
           shorten <=2pt,
           shorten >=2pt,},
    cascaded/.style = {%
    general shadow = {%
      shadow scale = 1,
      shadow xshift = -1ex,
      shadow yshift = 1ex,
      draw=black, very thick,
      text width=6.5em,
      fill = white},
    general shadow = {%
      shadow scale = 1,
      shadow xshift = -.5ex,
      shadow yshift = .5ex,
      draw=black, very thick,
           text width=6.5em,
      fill = white},
    fill = white, 
    draw=black, very thick,
    text width=6.5em,
    minimum width = 1.5cm,
    minimum height = 2cm}
}
\tikzstyle{arrow} = [thick,->,>=stealth]

\begin{document}
\title{The Sense of Logging in the Linux Kernel}
\subtitle{}


\author{Keyur Patel \and
    João Faccin \and
    Abdelwahab Hamou-Lhadj \and
    Ingrid Nunes 
}


\institute{K. Patel and A. Hamou-Lhadj \at
              Concordia University, Montreal, CA\\
              \email{\{keyur.patel,wahab.hamou-lhadj\}@concordia.ca}           
           \and
           J. Faccin and I. Nunes \at
              Universidade Federal do Rio Grande do Sul (UFRGS), Porto Alegre, BR\\
              \email{\{jgfaccin,ingridnunes\}@inf.ufrgs.br}
}

\date{Received: date / Accepted: date}

\maketitle


\begin{abstract}

Logging plays a crucial role in software engineering because it is key to perform various tasks including debugging, performance analysis, and detection of anomalies. Despite the importance of log data, the practice of logging still suffers from the lack of common guidelines and best practices. Recent studies investigated logging in C/C++ and Java open-source systems. In this paper, we complement these studies by conducting the first empirical study on logging practices in the Linux kernel, one of the most elaborate open-source development projects in the computer industry. We analyze 22 Linux releases with a focus on three main aspects: the pervasiveness of logging in Linux, the types of changes made to logging statements, and the rationale behind these changes. Our findings show that logging code accounts for 3.73\% of the total source code in the Linux kernel, distributed across 72.36\% of Linux files. We also found that the distribution of logging statements across Linux subsystems and their components vary significantly with no apparent reasons, suggesting that developers use different criteria when logging. In addition, we observed a slow decrease in the use of logging---reduction of 9.27\% between versions v4.3 and v5.3. The majority of changes in logging code are made to fix language issues, modify log levels, and upgrade logging code to use new logging libraries, with the overall goal of improving the precision and consistency of the log output. Many recommendations are derived from our findings such as the use of static analysis tools to detect log-related issues,  the adoption of common writing styles to improve the quality of log messages, the development of conventions to guide developers when selecting log levels, the establishment of review sessions to review logging code, and so on. Our recommendations can serve as a basis for developing logging guidelines as well as better logging processes, tools, and techniques.
\end{abstract}


\section{Introduction}\label{sec:intro}

Software logging is a practice that has long been used by developers to record valuable runtime information (e.g., events and data) about a running system~\citep{Yuan_2012a, pecchia2015industry, miranskyy2016operational}. These records are in the form of log messages, which are generated by logging statements in the source code. Studies have shown that the analysis of log data can help with many tasks, including  debugging and diagnosis of system failures~\citep{khatuya2018adele, el2020systematic}, detection of security vulnerabilities~\citep{yen2013beehive, zhou2020mobilogleak}, profiling of distributed systems~\citep{Pi_2018}, troubleshooting cloud computing environments~\citep{kc2011elt,miranskyy2016operational}, detection of anomalies~\citep{bertero2017experience, islam2018anomaly,oliner2008alert}, and reliability evaluation of  applications~\citep{tian2004evaluating}. 

Despite the importance of log analysis, the practice of logging remains largely \emph{ad-hoc} with no recognized guidelines and systematic processes~\citep{yuan2012characterizing, fu2014developers, pecchia2015industry}. Software developers continue to insert logging statements in the source code without clear and sufficient guidance. The decisions on how and where to log are often left to the discretion of the developers, resulting in inconsistencies even among developers working on the same projects. In a study performed at Microsoft~\citep{zhu2015learning}, the authors showed that 68\% of the participants (mainly Microsoft developers) found it challenging to make decisions on how and where to log. The lack of systematic and automated approaches for logging raises serious questions as to the validity of the generated log data as well as the efficacy of existing log analytic tools, which may impede failure diagnosis and other log-related tasks, threatening the stability of deployed software systems.



There exist studies that investigate the practice of logging in software engineering (e.g., ~\citep{zhou2020mobilogleak, yuan2012characterizing, chen2017characterizing, zhu2015learning,zeng2019studying}). \citet{yuan2012characterizing} made the first characterization of logging practices in open-source projects. The authors studied logging practices in four server-side C/C++ projects and provided many findings with respect to the pervasiveness of logging code, how often logging code is changed, and what kind of changes are made to it. \citet{chen2017characterizing} conducted an extensive replication study of this work with a focus on open-source projects written in Java. These studies revealed many issues related to the practice of logging. For example, they showed that software developers use logging in an inconsistent way. In addition, they showed that developers often have to change logging code in subsequent versions of the system as afterthoughts, adding to the maintenance burden of software systems. 

In this paper, we build on these studies by conducting the first empirical study to understand the practice of logging in the Linux kernel. Linux is perhaps one of the greatest collaborative efforts in the computer industry with more than a thousand experienced developers contributing to its growth on a regular basis. Many studies have examined the structure and evolution of the Linux kernel from different perspectives~\citep{bagherzadeh2018analyzing,israeli2010linux,lotufo2010evolution, passos_towards_2012,lu_study_2014, Fadel2011}, but little is known about the logging practices followed by  Linux kernel developers. Existing Linux development guidelines do not include guidelines for making logging decisions.

 Similarly to previous studies~\citep{yuan2012characterizing,chen2017characterizing}, we explore three main aspects related to the practice of logging: (1) the pervasiveness of logging in Linux, (2) the types of changes made to logging statements over several releases, and (3) the rationale underlying changes to logging code. From this perspective, this is a replication study with a focus on the Linux kernel. Because Linux developers are considered as experienced developers, understanding how they log their code provides useful insights into  the practice of logging in software engineering, further contributing to the corpus of knowledge in this area. Our findings show that there is a logging statement for every 27 lines of code in Linux v5.3 (the latest version when we collected data for this study). The pervasiveness of logging, however, varies from one kernel subsystem to another, with \textit{filesystem} and \textit{drivers} being the most logged subsystems. After investigating changes made to logging statements across 22 releases of Linux, we found that developers tend to modify logging code to improve the quality of the logging output by enhancing precision, conciseness, and consistency. By means of a qualitative analysis of 900 commits that target fixes and improvements of logging code, we found that the reasons behind these changes are similar to those reported by other studies~\citep{hassani2018studying, yuan2012characterizing,chen2019extracting,chen2017antipatterns}, such as ambiguous log messages, redundant information and logging the wrong variables. We also found many issues that pertain to the Linux kernel such as revealing vulnerable data and problems related to early logging. Based on these results, we derive many recommendations that are not only applicable to Linux, but can be readily generalized to other software systems. Our long-term goal is to help design better processes, techniques, and tools for effective logging, and set the groundwork for establishing logging standards.

In summary, the main contributions of our work are listed as follows.
\begin{itemize}
  \item We study the pervasiveness of logging in the Linux kernel. To our knowledge, this is the first study that focuses on the practice of logging in Linux.
   \item We present a study that shows the evolution of logging statements in 22 releases of the Linux kernel.
    \item We study the rationale behind changes made to logging statements across multiple releases of the kernel. 
    \item Based on the analysis of  afterthought changes to logging code in the Linux kernel, we derive several guidelines that can help the maintenance of logging code in Linux. These guidelines can possibly be applied to other systems as well. 
\end{itemize}

This study can benefit researchers and practitioners interested in understanding and improving the practice of logging in software engineering. The ultimate objective is to better define guidelines and processes for logging software systems. Additionally, Linux developers can directly refer to the findings of this paper to develop better logging practices for Linux.

The remainder of this paper is structured as follows. We discuss existing work related to logging in Section~\ref{sec:related-work}. Next, we present our study design in Section~\ref{sec:study-design}. The results obtained with the analysis of logging practices in the Linux kernel are presented in Section~\ref{sec:results}. In Section~\ref{sec:discussion}, we discuss our findings and implications as well as threats to validity. Finally, we conclude the paper in Section~\ref{sec:conclusion}.

\section{Related Work}\label{sec:related-work}

 In this section, we review existing studies on the practice of logging (i.e., how to log). We also review related studies on what and where to log. We discuss how these studies differ from ours at the end of the section.

\subsection{Research Studies on the Practice of Logging}

In an attempt to understand the characteristics of software logging in open-source projects,~\cite{yuan2012characterizing} conducted one of the first studies that examines  the practice of software logging by mining the evolution history of four C/C++ projects (Apache httpd, OpenSSH, PostgreSQL, and Squid). The authors found  that  logging statements account for 3.30\% of the source code. They also reported that developers often find it challenging to get logging statements right at the first try. They showed that developers changed 36\% of all log messages at least once as afterthoughts. The changes are categorized as changes to \emph{verbosity level}, \emph{static content}, and \emph{variables}. They observed that 75\% of log modifications of the static content of the logs consist of fixing inconsistent and confusing log messages. Additionally, the authors showed  that developers often struggle to use the right verbosity level. To address this problem, they designed a simple verbosity level checker. ~\citet{chen2017characterizing} conducted a replication study of the work of ~\citet{yuan2012characterizing} with the objective to understand the practice of logging in Java systems. Their  findings differ in many aspects from the original study.  For example, they found that consistency updates account for 41\% of all the logging code updates. These updates are much more present in C/C++ project and represent 67\% of the logging code updates.

\citet{pecchia2015industry} investigated the practice of logging on the development process of \emph{critical software systems} in industry. They identified three main reasons for using logging code: state dump, execution tracing, and event reporting. They found that the logging practice lacks standardization over key-value representations and missing contextual information, which hinders the adoption of automated log analysis tools.
 
\citet{shang2014} conducted a study on one large open-source and one industrial software system to explore the evolution of logging statements (referred to as communicated information) by mining the execution logs of these systems. They found that logging statements change across versions, impacting log processing applications. They also showed that the use of advanced log analysis techniques can reduce the impact of these changes on log preprocessing tools. 
 
\citet{zeng2019studying} focused on understanding the practice of logging in  \emph{Android apps}. Their findings suggest that there are differences between logging practices observed in mobile apps when compared to desktop and server applications. They provided evidence that logging code in mobile apps is less pervasive and not maintained as actively as in desktop and server applications.

The analysis of two open-source software systems conducted by~\citet{shang2015studying} showed a relationship between post-release defects and some log-related metrics. This study suggests that more maintenance effort should be devoted to files with logging code as they are more susceptible to contain problems. 

The quality of logging code was investigated by~\citet{hassani2018studying}. The authors characterized log-related issues using the number of files involved in an issue, the time required to fix it, and who provides the fix. In addition, they manually examined  log-related issues to identify common problems associated with logging code, which served as the basis for developing a tool to automatically detect problems such as spelling errors and empty catch blocks. 

\citet{cinque2010} conducted a quantitative study in which they assessed the effectiveness of logging mechanisms to generate logs when a fault is triggered. They injected faults in three open-source systems and observed whether faults that lead to failures generate logs  or not. The authors found that the percentage of logged failures ranges between 35.6\% and 42.1\% among the three systems they used in the case study. Their experimental results are then exploited to suggest places where to place additional logs based on the analysis of memory dumps.

Recently, \citet{li2020qualitativesurvey} conducted a qualitative study to understand the benefits of software logging and the costs associated with it by surveying 66 developers and examining 223 logging-related issue reports. The found that developers do not have a systematic  strategy for reconciling  the benefit and cost of logging. The authors argued that the lack of clear strategies may lead to logging practices with a negative impact.

 \citet{Li_2017a} proposed a predictive model that suggests the most appropriate log levels for new logging statements based on features such as the average log level and log churn. This is important in  light of the findings of ~\citep{yuan2012characterizing,chen2017characterizing,hassani2018studying} who showed that developers have difficulties  determining the correct log level to be used.

 \citet{chen2017antipatterns}  analyzed many code-independent log updates from three open-source Java projects and identified five logging anti-patterns. A tool, called \emph{LCAnalyzer}, was developed to allow the automatic identification of these poor logging practices. 

\citet{zhou2020mobilogleak}  investigated the impact of logging practices on data leakage with a focus on Android mobile applications. They studied logging code in 5,000 mobile apps and showed using taint analysis that out of 276 apps with taint flow paths, 200 (72\%) leaked sensitive data due to poor logging
practices such as forgetting to remove debug logging statements before deployment. They categorized the types of data leakage that are related to logging practices into four types: network, account, location, database. They manually analyzed source-sing paths of log-related taint flow paths of the 200 apps and found that 49\% of the data that is leaked is network-related, followed by database sensitive data (45\%).

\subsection{Studies on where and what to log}

\citet{fu2014developers} performed an empirical investigation on enterprise applications written in C\# in order to understand where developers insert logging statements. The authors found that logging statements can be grouped based on where they are inserted in the code using these categories: \emph{assertion-check logging}, \emph{return-value-check logging}, \emph{exception logging}, \emph{logic-branch logging}, and \emph{observing-point logging}. They also found that 39\%--53\% of logging statements are placed to capture information when the software fails while 47\%--61\% of logging statements are used to trace the normal execution flow. To complement their findings, the authors trained a decision tree model that suggests where to log based on contextual keywords extracted from code snippets. Their recommendation tool achieved a precision of 81.1\%--90.2\%, and recall of 80.8\%--90.4\%. A similar tool, called \emph{LogAdvisor}, was proposed by \citet{zhu2015learning}. The tool aims to help developers by providing suggestions on where to log with an accuracy of 84.6\%--93.4\%. 
 
In their position paper, \citet{xu2017game} proposed a strategy for automatically placing logs in the source code using information theory concepts by measuring the software’s uncertainty using Shannon Entropy. Their approach respects a predefined performance overhead  (less than 2\%). The idea is to place logs in execution paths that reduce uncertainty. The approach was tested on a small example. The authors stated that the preliminary results show that their approach holds real promise for automating the placement of logging statements. The problem with this technique, however, is that it is completely agnostic to the semantic of the code blocks that are being logged. It simply adds logs in paths that would most likely be executed. The authors did not discuss the quality of the generated logging statements when used for debugging and other tasks.

\citet{Yang2018nano} proposed NanoLog, a logging mechanism that is 1--2 orders of magnitude faster than existing logging systems and that achieves a high throughout (80 million log messages per second), while exposing the traditional printf-like statements. NanoLog modifies the log messages at compile-time, and uses compaction techniques to output logs in a compact binary format. The authors showed that their approach can improve the performance of log analytic tools.

Because there is no standard way to know what information should be included in logs, developers insert logging statements into the code in an \textit{ad-hoc} manner. Sometimes, the information from these logs may be insufficient to diagnose software failures. To handle this issue, ~\citet{Yuan_2012a} designed \emph{LogEnhancer} a tool that can infer  \emph{what} information could be helpful to narrow down the root cause of failures and to include that information into existing logging code automatically. To help developers decide which variables should be included in the logging statements, a recent study by~\citet{varstolog2019} proposed a recurrent neural network-based model. The work of \citet{He_2018a} also focuses on what to log, with the objective to  understand the goal of the static text used in these statements. Their results suggest that static text is used essentially for describing program operation, error conditions, and high-level code semantics. They also proposed an information retrieval approach for generating logging messages automatically. 

\citet{li2018studying} used topic modelling to understand whether there is any relationship between \textit{topics} of source code and  having logging statements. Their findings suggest that some source code \textit{topics} are more likely to include logging statements than others.

The single attempt to address the challenge of logging in Linux was made by \citet{tschudin20153l}. The authors proposed a machine learning-based approach to suggest the most appropriate logging functions, using the evolution history of the Linux kernel. However, the approach consists of a conceptual description of a framework that still needs to be implemented and validated.

\subsection{Discussion}

Our study focuses on understanding the practice of logging adopted by Linux developers. To our knowledge, this is the first study that focuses on the Linux kernel, which makes our study unique. To achieve our goal, we address the same research questions on characterizing the practice of logging as the ones discussed in related work ~\citep{yuan2012characterizing,chen2017characterizing}. In this sense, this study can be considered as a replication study with a focus on the Linux kernel. It complements the studies that aim to better understand of the practice of logging in software development. We found that Linux developers tend to modify logging code to improve the quality of the logging output by enhancing precision, conciseness, and consistency. We also found that the reasons behind afterthought changes to logging statements include to the ones reported in  other studies~\citep{hassani2018studying, yuan2012characterizing,chen2019extracting,chen2017antipatterns}. This confirms the need for better guidelines and processes for logging in software engineering. We derive many recommendations from our study that can improve the logging practices in Linux and other systems. Many of these can be implemented in tools such as the ones discussed in the previous subsection to make suggestions to developers as they write logging code.

\section{Study Design}\label{sec:study-design}

\subsection{Research questions}

Our goal is to characterize how logging is used in the Linux kernel so that we can understand the adopted practices and identify the challenges. To achieve this goal, we focus on answering the following research questions.

\begin{description}
    \item[RQ1] \textit{What is the pervasiveness of logging in the Linux kernel?}
    \item[RQ2] \textit{How does the logging code in the Linux kernel evolve?}
    \item[RQ3] \textit{What are the causes other than keeping consistency in the code that lead developers to change the logging code?}
\end{description}

By answering RQ1, we aim to understand whether logging is a practice consistently adopted by Linux kernel developers. This involves the analysis of the location of logging statements to understand the rationale of why they are placed in specific locations. In RQ2, we focus on investigating the lifecycle of logging statements, i.e.\ when they are created, updated, or removed. By making such analyses, we are able to understand the state of practice of the Linux kernel, that is, how much logging exists in the code, how it is distributed, and its lifestyle. The goal is to observe discrepancies in how logging is adopted so that we can raise questions related to the adopted practices. Such questions can be the basis for future studies to better understand logging decisions or to improve how logging is added and maintained in the code.

These two research questions are similar to those answered in previous studies~\citep{yuan2012characterizing,chen2017characterizing}. This would allow us not only to understand these aspects in a project with distinguished characteristics but also to contrast obtained results. These previous studies also further analyzed logging changes to logging code, classifying them as consistent changes and afterthought changes. The former refers to changes that are due to changes in other parts of the code that requires the logging code to be updated to keep consistency, for example, when a variable referred to in a logging statement has its name changed. Afterthought  changes, in turn, are changes made to logging statements to fix or improve the logging statements for other purposes. In this paper, we investigate such purposes in RQ3 to provide insights with respect to the effort required to maintain logging code. We aim to identify the reasons behind afterthought changes, which we hope can help researchers and practitioners develop techniques to automatically prevent, detect and automate such changes.

\subsection{Subject project}\label{subsec:subject}

The Linux kernel, our subject project, is a free and open-source software distributed under the GPLv2\footnote{\url{https://www.gnu.org/licenses/old-licenses/gpl-2.0.html}} license. It is responsible for managing the interactions between hardware components and higher-level programs that use these components. Used across a wide range of computer systems, from mobile devices and server systems to supercomputers, the kernel can be considered as one of the most important software projects in the computing industry. In fact, we are surrounded by the Linux kernel in one way or another. Being developed by roughly fifteen thousand developers throughout its history, as of 2020, it contains around 18M lines of C code. This project is continuously evolving to meet both hardware manufacturers' requirements and end-user expectations. 

For RQ1, we study the pervasiveness of logging in version v5.3 (the last version when doing this study). This version received 14,605 commits from 1,881 developers, adding 837,732 and removing 253,255 lines of code, which represents an increase of 584,477 lines of code when compared to its previous version.\footnote{\url{https://github.com/gregkh/kernel-history/}}

To understand changes to logging code across various Linux releases and hence answer RQ2 and RQ3, we need to select a number of releases to include in the study. Our first attempt was to cover ten years of development going back starting from v5.3 (the last version when we conducted this study), but the amount of data that was generated was simply too high to conduct a viable analysis, especially that there is a lot of manual work to answer RQ3 and part of RQ2. We therefore decided to focus on a smaller dataset without loss of generality by going back 20 releases starting from v5.3. We settled for v4.3\footnote{\url{https://github.com/torvalds/linux/commit/6a13feb9}} to v5.3\footnote{\url{https://github.com/torvalds/linux/commit/4d856f72}} (22 releases to be exact). This corresponds to an interval of  four years of software development, from November 2015 to September 2019. All releases used in this study are available in the official Linux kernel repository.\footnote{\url{https://github.com/torvalds/linux}} We believe that this dataset is  representative since our objective is not to uncover all the problems related to logging in Linux but rather to provide insight into the practice of logging in software development by looking at how Linux developers (even in a narrower scope) use logging. 

\subsubsection{Linux subsystems}

In our analysis, we frequently refer to  Linux subsystems. The Linux kernel consists of five major subsystems;\footnote{\url{https://github.com/gregkh/kernel-history/blob/master/scripts/genstat.pl}} each covering particular aspects of the project. Table~\ref{tbl:subsystems} shows the system decomposition. The \emph{core} subsystem is in charge of memory management, inter-process communication, management of I/O operations, among others. The \emph{filesystem} subsystem is responsible for providing the file system interface as well as individual file system implementations. The \emph{drivers} subsystem provides device and sound drivers as well as the implementation of cryptography algorithms used within the system and other security-related code. Finally, the \emph{net} and \emph{arch} subsystems are responsible for networking and architecture-specific code, respectively.

\begin{table}
 \caption{Linux kernel architectural decomposition.}\label{tbl:subsystems}
 \centering
 \begin{tabular}{ll}
 \toprule
 \textbf{Subsystem} & \textbf{Top level directories} \\ \midrule
 \emph{core} & init, block, ipc, kernel, lib, mm, virt \\
 \emph{filesystem} & fs \\
 \emph{drivers} & crypto, drivers, sound, security \\
 \emph{net} & net \\
 \emph{arch} & arch \\ \bottomrule
 \end{tabular}
\end{table}

\subsubsection{Logging in the Linux kernel}

Logging statements are used to record relevant information about a running system. A logging statement usually comprises three elements: (i) the \emph{log level} of the event being recorded; (ii) a \emph{static message} that describes that event; and, optionally, (iii) values of \emph{variables} related to the logged event. A log message is typically generated by the execution of a function specifically created for that purpose. These functions can be either project-specific or provided by external logging libraries and frameworks.

The Linux kernel provides developers with its own set of logging functions. One of the simplest ways to write a message to the kernel log buffer is by using the \texttt{printk()} function. It is the kernel's equivalent of \texttt{printf()}, with the difference that it allows developers to specify the log level of the event being recorded~\citep{corbet2005linux}. An example of a logging statement is as follows:

\begin{verbatim}
  printk(KERN_ERR "Device initialized with return code %d\n", code);
\end{verbatim}

\noindent where \texttt{KERN\_ERR} corresponds to the log level, \texttt{"Device initialized with return code \%d\textbackslash n"} is the static message, and \texttt{code} is the corresponding variable. There are eight log levels defined in the Linux kernel, representing different degrees of severity (see  Table~\ref{tbl:loglevels}). \texttt{KERN\_WARNING} is the default log level and is assigned to a message in case no log level is specified when calling \texttt{printk()}.

\begin{table}
    \caption{The eight possible log levels defined in the Linux kernel.} \label{tbl:loglevels}
    \centering
    \begin{tabular}{ll}
        \toprule
        \textbf{Log level} & \textbf{Description} \\ \midrule
         \texttt{KERN\_EMERG} & System is unusable \\
         \texttt{KERN\_ALERT} & An action must be taken immediately \\
         \texttt{KERN\_CRIT} & Critical conditions \\
         \texttt{KERN\_ERR} & Error conditions \\
         \texttt{KERN\_WARNING} & Warning conditions (default) \\
         \texttt{KERN\_NOTICE} & Normal but significant condition \\
         \texttt{KERN\_INFO} & Informational \\
         \texttt{KERN\_DEBUG} & Debug-level messages \\ \bottomrule
    \end{tabular}
\end{table}

Additional sets of logging functions were introduced in the Linux kernel v1.3.983 with the aim of making logging statements more concise. These functions incorporate log levels in their names. Therefore, in order to log a debug or informational message, instead of using the \texttt{printk()} function with the \texttt{KERN\_DEBUG} and \texttt{KERN\_INFO} levels as parameters, developers can use the \texttt{pr\_debug()} and \texttt{pr\_info()} functions, respectively. Another family of functions specifically designed for device drivers, e.g.\ \texttt{dev\_dbg()} and \texttt{dev\_info()}, automatically include device names in their outputs, making it easier to identify the origin of log messages. Table~\ref{tbl:printk-helpers} lists both sets of functions and their corresponding log levels. Currently, some kernel components present their own logging functions, which are able to generate messages with service-specific information. Examples include network device and TI wl1251 drivers, which provide the \texttt{netdev\_*()} and \texttt{wl1251\_*()} families of logging functions, respectively.

\begin{table}
    \caption{New sets of logging functions.}
    \label{tbl:printk-helpers}
    \centering
    \begin{tabular}{lll}
        \toprule
        \textbf{Log level} & \textbf{\texttt{pr\_*()} function} & \textbf{\texttt{dev\_*()} function} \\ \midrule
        \texttt{KERN\_EMERG} & \texttt{pr\_emerg()} & \texttt{dev\_emerg()} \\
         \texttt{KERN\_ALERT} & \texttt{pr\_alert()} & \texttt{dev\_alert()} \\
         \texttt{KERN\_CRIT} & \texttt{pr\_crit()} & \texttt{dev\_crit()} \\
         \texttt{KERN\_ERR} & \texttt{pr\_err()} & \texttt{dev\_err()} \\
         \texttt{KERN\_WARNING} & \texttt{pr\_warn()} & \texttt{dev\_warn()} \\
         \texttt{KERN\_NOTICE} & \texttt{pr\_notice()} & \texttt{dev\_notice()} \\
         \texttt{KERN\_INFO} & \texttt{pr\_info()} & \texttt{dev\_info()} \\
         \texttt{KERN\_DEBUG} & \texttt{pr\_debug()} & \texttt{dev\_dbg()} \\ \bottomrule
    \end{tabular}
\end{table}

\subsection{Procedure}\label{procedure}

\begin{figure}
    \centering
    \includegraphics[width=\linewidth]{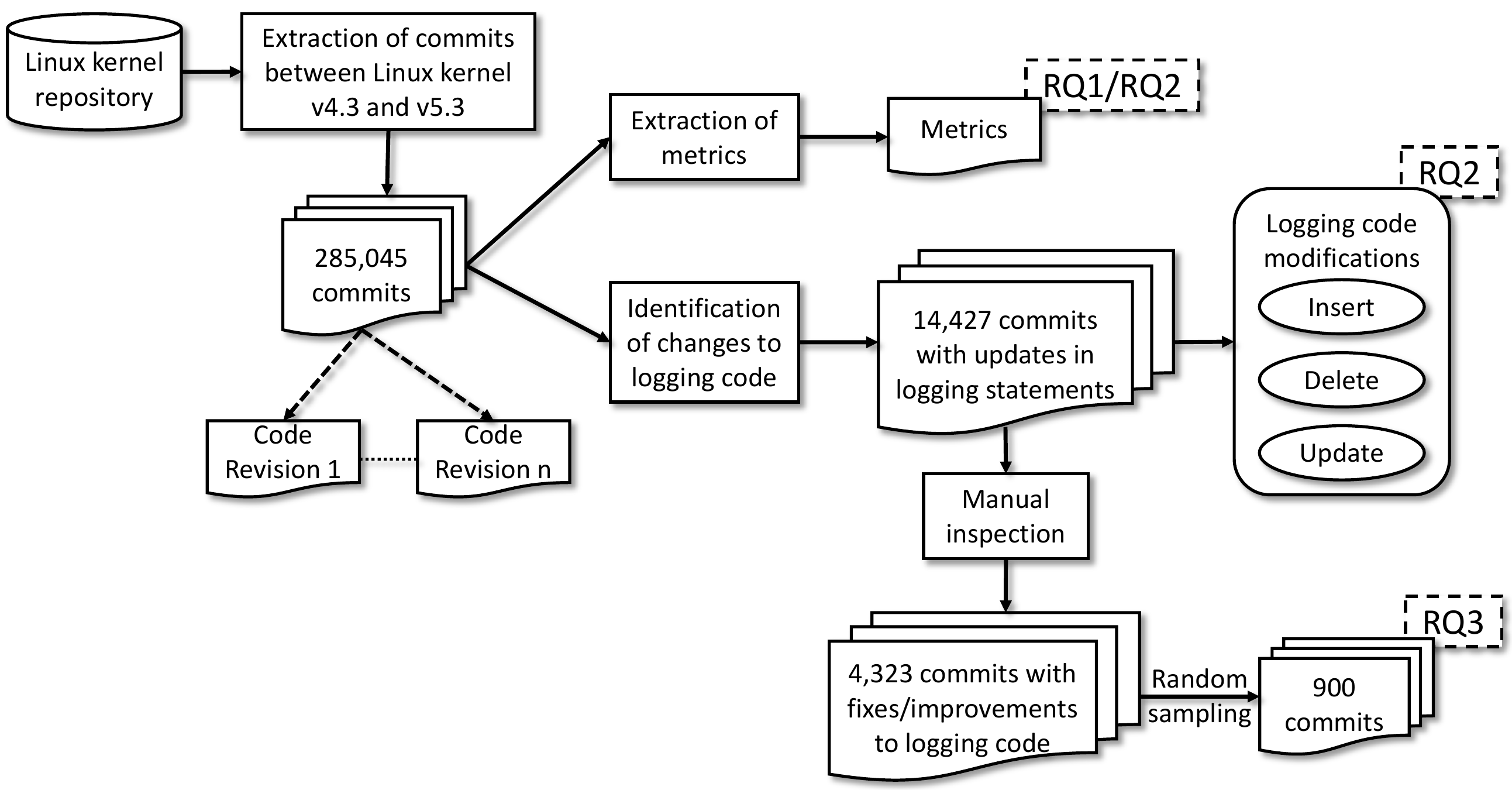}
     \caption{Overview of the study procedure.}
    \label{fig:procedure}
\end{figure}

In this section, we detail how we collected logging data from the Linux kernel. Figure~\ref{fig:procedure} summarizes the tasks that were carried out. In the following subsections, we describe how we (1) extracted logging statements from the target project; (2) calculated metrics; (3) identified changes made to logging code between different versions; and (4) selected a subset of changes in logging statements to be analyzed.

\subsubsection{Identification of logging functions}\label{subsec:identification}

Traditional approaches to automatically locate logging statements into the source code rely on logging functions that are known in advance (e.g., Log4j in Java) or use regular expressions looking for variations of the term ``log''. Although these methods have been successfully adopted in previous studies \citep{shang2015studying, zhu2015learning, chen2017characterizing,yuan2012characterizing}, they are not applicable in the context of the Linux kernel, which uses a wide variety of functions and macros for logging purposes. It is common to find situations in which customized calls to these macros are implemented by particular services (e.g., \spverb|_enter|\footnote{\url{https://github.com/torvalds/linux/blob/v5.3/fs/afs/internal.h\#L1449}}, \spverb|pr_pic_unimpl|\footnote{\url{https://github.com/torvalds/linux/blob/v5.3/arch/x86/kvm/i8259.c\#L37}}). Listing~\ref{lst:macro} shows an example. A macro named \texttt{adsp\_dbg()} is implemented with the aim of ensuring the invocation of \texttt{dev\_dbg()} with a particular set of parameters and formatting (lines 2--3) so that there is no need to specify such elements in further calls (line 5). This type of logging statements cannot be detected unless we know that \texttt{adsp\_dbg()} is a logging function.

\noindent\begin{minipage}[t]{0.95\textwidth}
\begin{lstlisting}[language=C, caption={Logging macro defined in /sound/soc/codecs/wm\_adsp.c.}, label=lst:macro, xleftmargin=20pt, numbers=left, captionpos=b]
[...]
#define adsp_dbg(_dsp, fmt, ...) \
	dev_dbg(_dsp->dev, "%s: " fmt, _dsp->name, ##__VA_ARGS__)
[...]
adsp_dbg(dsp, "Wrote %zu bytes to %x\n", len, reg);
[...]
\end{lstlisting}
\end{minipage}

To identify Linux logging functions and macros  in our study, we take advantage of the pattern-based method presented by~\citet{tschudin20153l}. The authors proposed an identification method that consists of three semantic patterns. These patterns, although describing general properties when taken individually, are able to correctly characterize logging functions when combined, yielding a low number of false positives. The first pattern states that logging functions should be called at least once inside an \texttt{if} block that ends with a \texttt{return} statement. The second pattern indicates that logging functions have at least one string argument, representing a log message. Finally, the third pattern requires logging functions to have a variable number of arguments. Listing~\ref{lst:pattern} shows how macro \texttt{adsp\_dbg()} satisfies all three patterns and thus can be considered as a logging function. 

\noindent\begin{minipage}[t]{0.95\textwidth}
\begin{lstlisting}[language=C, caption={A logging macro defined in /sound/soc/codecs/wm\_adsp.c that satisfies the identification patterns.}, label=lst:pattern, xleftmargin=20pt, numbers=left, captionpos=b]
[...]
if (val == 0) {
    adsp_dbg(dsp, "Acked control ACKED at poll %u\n", i);
    return 0;
}
[...]
adsp_dbg(dsp, "Wrote %zu bytes to %x\n", len, reg);
\end{lstlisting}
\end{minipage}

Although this approach yields good results, it might miss logging functions with a fixed number of arguments (e.g., \spverb|PRINTK_2|\footnote{\url{https://github.com/torvalds/linux/blob/v5.3/drivers/char/mwave/mwavedd.h\#L79}} and \spverb|PRINTK_4|\footnote{\url{https://github.com/torvalds/linux/blob/v5.3/drivers/char/mwave/mwavedd.h\#L89}}). These functions do not satisfy the third pattern. We therefore decided to relax the third pattern by including logging functions that contain a fixed number of arguments (i.e., the ones that are not overloaded).  

We have carefully reviewed the list of logging functions extracted with this approach. Although we believe that this is a very comprehensive list of functions, we cannot guarantee 100\% coverage, which is a threat to internal validity of this study. To mitigate this threat, we selected randomly 20 files and checked manually their content to see if we  missed any logging functions. We found that our list  covered all the logging functions invoked in these files. Further, we make the list of logging functions (and the data used for RQ1, RQ2, and RQ3) available to the community.\footnote{\url{https://github.com/iamkeyur/linux-logging-2021}} Once we know which functions/macros are used for logging, we write a script to detect all calls to these functions/macros to retrieve the specific logging statements. 

Another alternative to handle the problem of the many forms of log printing functions caused by macros would be to expand macro definitions using a compiler preprocessor (e.g., gcc -E). However,  ~\citet{tschudin20153l} found that this approach may result in false positives because of the possibility of having non-logging functions that contain calls to printk-like functions.

\subsubsection{Logging code metrics}

To assess the pervasiveness of logging in the Linux kernel, we consider four metrics, summarized in Table~\ref{tbl:rq1:metrics}. Similarly to previous studies (e.g., \cite{yuan2012characterizing}), we extract the number of \emph{lines of source code} (SLOC) and the number of \emph{lines of logging code} (LLOC). We only included \texttt{.c} files in our analysis, that is, header files were not taken into account. Comments and empty lines are discarded when measuring SLOC. These metrics are collected considering different \emph{levels} of granularity: (i) overall system and subsystems; (ii) files; and (iii) program constructs. The latter comprises functions as well as blocks of control statements, namely \texttt{do-while}, \texttt{if}, \texttt{else}, \texttt{else-if}, \texttt{for}, \texttt{switch}, and \texttt{while}. 
We also compute the \emph{Log Density}~\citep{yuan2012characterizing, chen2017characterizing, zeng2019studying} and \emph{Log Ratio} metrics (see Equations~\ref{eq:density} and~\ref{eq:ratio}). The Log Density measures the number of lines of source code per logging statement, whereas the Log Ratio measures the number of logging statements per line of source code. For example, a log density of 10 would mean that for every ten lines of code, there is one logging statement. A log ratio of 0.2 means that 20\% of the code consists of logging statements.
\begin{multicols}{2}
\begin{equation}\label{eq:density}
Log~Density = \frac{SLOC}{LLOC}
\end{equation}\break
\begin{equation}\label{eq:ratio}
Log~Ratio = \frac{LLOC}{SLOC}
\end{equation}
\end{multicols}

\begin{table}
\small\centering
\caption{Extracted metrics related to logging code.}
\label{tbl:rq1:metrics}
\begin{tabular}{ll}
\toprule
\textbf{Metric} & \textbf{Description}\\
\midrule
SLOC & Number of lines of source code excluding comments and empty lines\\
LLOC & Number of lines of logging code\\
Log Density &  The number of lines of source code per logging statement\\
Log Ratio & The percentage of number of lines of logging code per source line of code\\
\bottomrule
\end{tabular}
\end{table}

\subsubsection{Detection of changes in logging statements} 

To understand how logging code evolves in the Linux kernel, we study three types of modifications made to logging code---log insertion, log deletion, and log update---similarly to previous work~\citep{yuan2012characterizing,chen2017characterizing,zeng2019studying}. We further divide insertion and deletion of logging statements into two categories: logging statements that are added or deleted along with the addition or deletion of files. To retrieve changes to logging statements, we wrote a script using the \texttt{git rev-list} command with \texttt{--no-merges} option to retrieve the commits. Note that the \texttt{--no-merges} option omits commits with more than one parent. This is necessary in order to avoid duplicate commits. For each commit, we extract two adjacent versions of each file that are changed in the commit. We only consider \texttt{.c} files. Then, to generate an edit script representing syntactic modifications by inferring changes at the level of abstract syntax tree, we use GumTree, which is the state-of-the-art AST differencing tool~\citep{FalleriMBMM14}. As result, we found that 14,427 commits modified at least one logging statement, which are those used to answer RQ2.




\subsubsection{Selection of changes in logging statements} 

Considering the 14,427 commits used to answer RQ2, not all aim to provide fixes or improvements to logging code, which is the type of change in logging code that we are interested in to answer RQ3. To identify changes that were made targeting logging code, previous studies~\citep{hassani2018studying,mazuera2020investigating} rely on a keyword-based approach by searching for commits that contain in their message variations of the word ``log''. This strategy may miss many commits in Linux such as the one described above. To address this, the first author of this study manually examined all 14,427 commits by reading the commit titles and messages. He identified 4,323 commits out of 14,427 commits that explicitly discuss changes to logging statements. This task took several weeks to complete and required multiple iterations. He then selected randomly 900 commits out of 4,323 for a qualitative analysis. This subset is statistically representative with a confidence level of 99\% and a margin of error of $\pm$ 4\%~\citep{boslaugh2012statistics}. 


\section{Results and Analysis} \label{sec:results}

We followed the previously described study procedure and obtained the results detailed in this section. We present and analyze the results by research question.


\subsection{RQ1: Pervasiveness of logging in Linux kernel}\label{sec:rq1}

To examine the pervasiveness of logging in the Linux kernel, we consider its latest available version when we collected data for this study (v5.3). Based on the introduced metrics, we quantitatively assess how logging statements are spread in the code, considering four granularity levels: system (i.e., the entire Linux kernel), subsystems, files, and programming constructs.


\subsubsection{System and subsystem level}

We first report the values obtained for each metric (SLOC, LLOC, Log Density and Log Ratio) considering the system and subsystems levels. Table~\ref{tbl:rq1:results} presents the results. We found that from a total of 13,390,104 lines of source code, 498,897 (i.e., 3.73\%) are lines of logging code. It is equivalent to a logging density of 27, which means that for every 27 lines of source code in Linux there is one logging statement. This finding is in line with the observations made by~\citet{yuan2012characterizing} who reported an average log density of 30 in the four C/C++ applications they examined. This result, however, differs from the study by~\citet{chen2017characterizing} who reported an average log density of 50 in  the 21 Java applications they studied. Although these results are not conclusive, they suggest that the language  in which the project is written \textcolor{black}{as well as the application domain} may affect the prevalence of logging statements in the source code. \textcolor{black}{In addition, this gives evidence of the relevance of logging code. Logging statements do not provide additional application functionality and they, still, account for almost 4\% of the written code in our target system. Consequently, logging code should be maintained with the same practices of code associated with other concerns. For example, developers should carefully review logging code in code review.}

\begin{table}
    \caption{Summary of metrics collected from the Linux kernel and its subsystems.}
    \label{tbl:rq1:results}  
    \centering \small
    \begin{tabular}{llrrrr}
        \toprule
        \textbf{Subsystem} & \textbf{Component} & \textbf{SLOC} & \textbf{LLOC} & \textbf{Log Density} & \textbf{Log Ratio}\\ \midrule
        \multirow{8}{*}{core} & lib & 106,577 & 2,296 & 46 & 2.15\%\\ 
        & kernel & 208,000 & 4,396 & 47 & 2.11\%\\ 
        & mm & 86,921 & 1,940 & 45 & 2.23\%\\
        & block & 31,846 & 713 & 45 & 2.24\%\\ 
        & ipc & 6,468 & 13 & 498 & 0.20\%\\ 
        & init & 2,935 & 192 & 15 & 6.54\%\\
        & virt & 16,839 & 184 & 92 & 1.09\%\\ \cmidrule(lr){3-6}
        & & 459,586 & 9,734 & 47 & 2.12\%\\ \midrule
        filesystem & fs & 840,527 & 35,780 & 23 & 4.26\%\\ \midrule
        \multirow{4}{*}{drivers} & drivers & 9,358,913 & 384,269 & 24 & 4.11\%\\
        & sound & 765,988 & 19,673 & 39 & 2.57\%\\
        & security & 57,974 & 1,192 & 49 & 2.06\%\\
        & crypto & 52,888 & 835 & 63 &  1.58\%\\ \cmidrule(lr){3-6}
        & & 10,235,763 & 405,969 & 25 & 3.97\%\\ \midrule
        net & net & 746,263 & 14,490 & 52 & 1.94\%\\ \midrule
        arch & arch & 1,107,992 & 32,924 & 34 & 2.97\%\\ \midrule
        \textbf{Total} & & \textbf{13,390,131} & \textbf{498,897} & \textbf{27} & \textbf{3.73\%}\\
    \bottomrule
    \end{tabular}
\end{table}

When looking at each subsystem individually, we obtain heterogeneous results.  The log ratio ranges from 1.94\% to 4.26\%. A deeper look at the components of the Linux subsystems shows important discrepancies. For example, we observe that the \textit{init} component of \textit{core}, which consists only of 2,935 SLOC, contains 192 lines of logging code (Log Ratio = 6.54\%). This is considerably higher than the other components. This may be explained by the fact the \textit{init} component is responsible for the initialization of the console and other key kernel services such as the security framework, scheduler, memory allocation~\citep{2020bootlin}. For these services, it is important to log all possible errors in order to quickly debug potential failures. However, the idea that critical components are logged the most does not always hold. Take for example the \textit{ipc} component from the \textit{core} subsystem with the highest log density (Log Density = 498). This component, which contains only six files, is responsible for setting up the inter-process communication mechanisms on which the kernel processes rely. Yet, despite being critical, this component is the least logged.

Without further studies, we can only attribute these variations to the fact that different groups of developers are maintaining each subsystem, and there are no recognized (or common) guidelines on how to log. \textcolor{black}{This calls for future studies to understand the rationale for adding a large amount of logging statements in particular modules of the code considering different goals, such as debugging or auditing. It is important to understand when it is critical to have detailed logged information, e.g.\ when a module have a specific role or is the target of frequent changes. Moreover, given that logging can have a performance impact, certain components may have a reduced number of logging statements due to that reason.}

\subsubsection{File level}

\begin{figure}
    \centering
    \includegraphics[width=\linewidth]{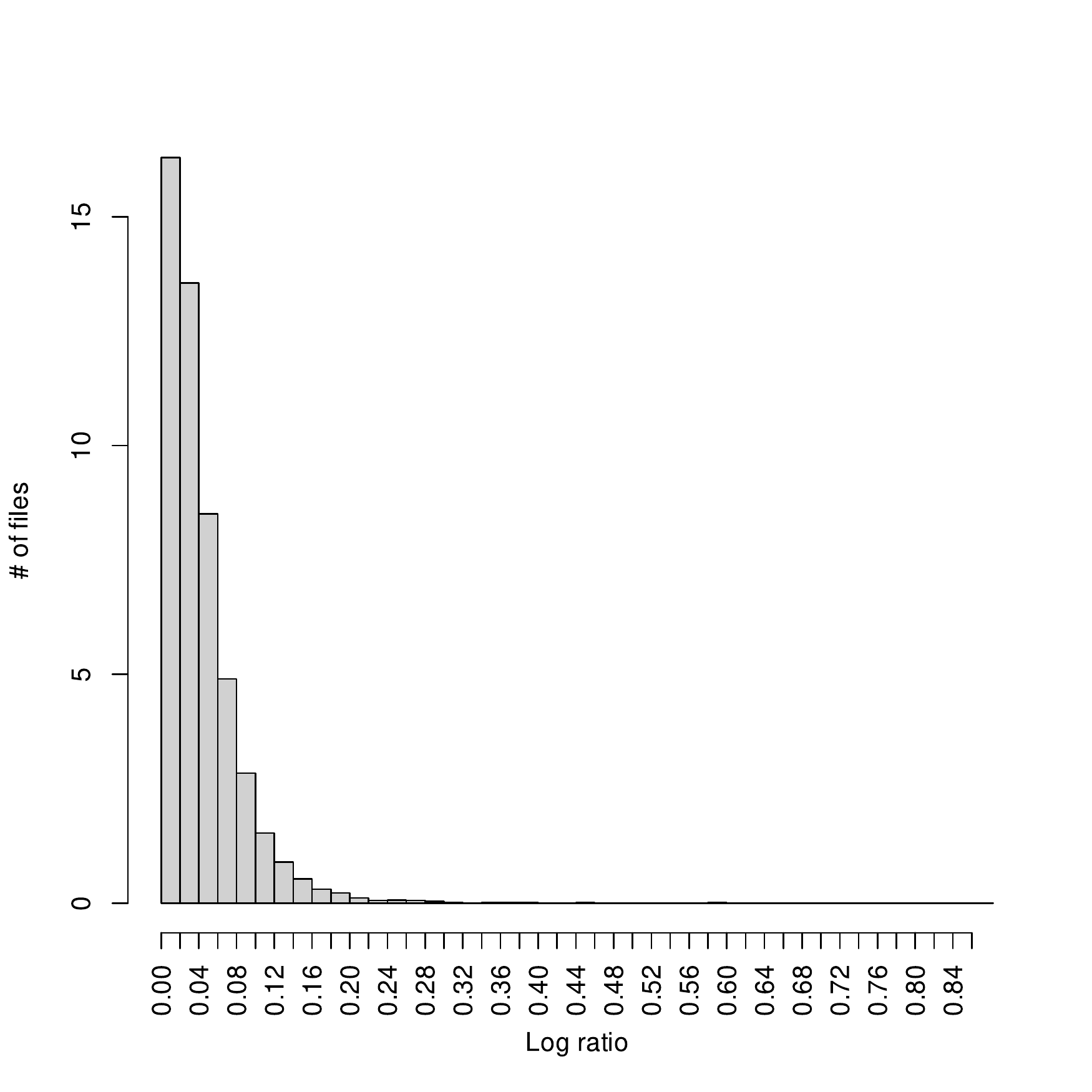}
    \caption{Value of log ratio at the file level (Total \#files = 25,814).}
    \label{fig:rq1}
\end{figure}

Similarly to the study of \cite{lal2015two}, we also calculate the log ratio at the file level. Note that we do not use log density here due to the fact that many files do not contain logging statements, and therefore this metric cannot be calculated because SLOC cannot be divided by zero. The log ratio provides a more accurate measurement by assessing the pervasiveness of logging across Linux files. Figure~\ref{fig:rq1} shows the collected data. We found that 94.34\% of the total number of files in Linux v5.3 have a log ratio between 0\% and 10\%. More precisely, 7,134 files (27.64\%) do not contain any logging statements, while a significant number of files (66.70\%) have a log ratio greater than 0\% and less than or equal to 10\%. Files with a log ratio greater than 10\% account for only 5.66\% of the total number of files. \textcolor{black}{Figure~\ref{fig:rq1} thus highlights that the number of logging statements by file is a skewed distribution. Having the majority of files with a few logging statements is expected because many files may include code statements that can result in errors, and these are typically logged. However, it is interesting to observe that there are files with a substantial amount of logging statements. As discussed above, a qualitative study focusing on these highly logged files could give directions why this occurs and is needed. We, in particular, manually inspected} 6 files with a log ratio close to 90\%. We found that they contain mostly debugging routines. For example, \path{drivers/scsi/qla4xxx/ql4_dbg.c} contains functions to dump relevant information about the Linux Host Adapter structure.

\begin{figure}
    \centering
    \includegraphics[width=\linewidth]{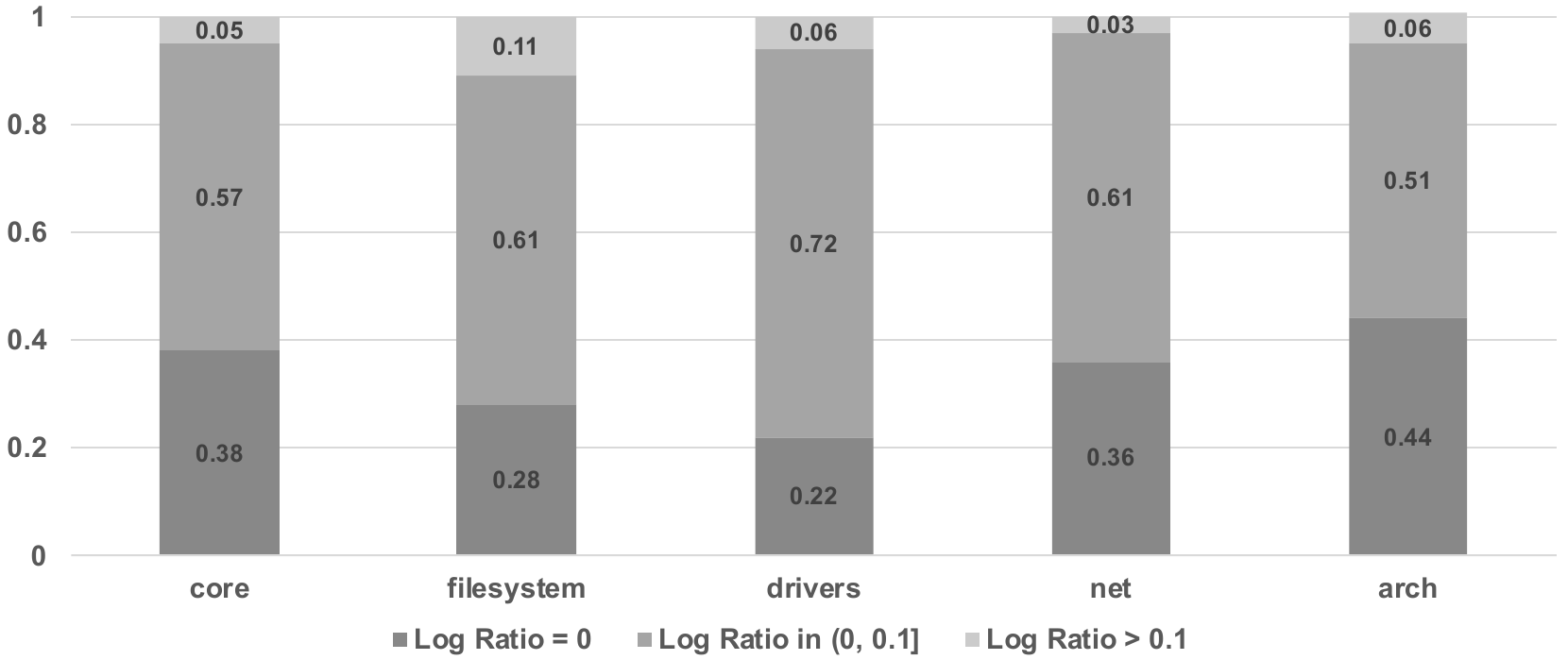}
     \caption{Distribution of files by subsystem based on the log ratio.}
    \label{fig:logDensityBySubsytem}
\end{figure}

The analysis of this distribution by the Linux subsystems can be seen in Figure~\ref{fig:logDensityBySubsytem}, which shows the distribution of files by subsystem according to their log ratio. We can observe that 22\%--44\% of the files have no logging statements in the subsystem files. This indicates that in some systems---\textit{core}, \textit{net}, and mainly \textit{arch}---the logging is concentrated in fewer files. Except \textit{filesystem}, only a small percentage of files (3\%--5\%) has a high log ratio, that is, having many logging statements is an exception. This is not the case of \textit{filesystem}, which has 11\% of the files with more than 10\% of their code being logging statements. In all subsystems, the majority (sometimes nearly half) of the files have at most 10\% of their code as logging statements, but contain at least one. Similarly to what was discussed above, these differences cannot be explained without future studies. 

\subsubsection{Programming construct level}

In this section, we measure the number of  logging statements in program constructs of the Linux kernel including functions, \texttt{do-while}, \texttt{if}, \texttt{else}, \texttt{else-if}, \texttt{for}, \texttt{switch}, and \texttt{while}. Determining where logging statements are located can help understand the purpose they serve. There exist studies that examine the location of logging statements in the source code. In particular, the work by~\citet{pecchia2015industry} measured the number of logging statements in the same program constructs as those listed above. Their work targeted two large C/C++ industrial applications. We followed the same approach and compared our results to theirs.

Table~\ref{tbl:logpermethod} shows the distribution of logging statements in the kernel functions. We found that from a total of 476,522 functions, 352,045 (73.88\%) do not have any logging statements. For the remaining functions (26.12\%), 88.28\% of these have the number of logging statements in the range of one to five. Less than 1\% of the functions have more than eight logging statements. \textcolor{black}{Here, again, the distribution of logging statements is skewed. But differently, we observe that the vast majority of functions do not contain any logging statements at all. This shows that a relatively low number of functions need to have their information recorded in logs.}

\begin{table}
\caption{Number of functions that have a certain amount of logging statements.}
\label{tbl:logpermethod}
\centering
\begin{tabular}{lr}
\toprule
\#Logging Statements &  \#Functions\\ 
\midrule
0 & 352,045 (73.88\%) \\
1 & 64,553 (13.55\%) \\ 
2 & 25,823 (5.42\%) \\ 
3 & 12,432 (2.61\%) \\ 
4 & 7,075 (1.48\%) \\ 
5 & 4,432 (0.93\%) \\ 
6 & 2,752 (0.58\%) \\ 
7 & 1,902 (0.40\%) \\ 
8 & 1,282 (0.27\%) \\ 
$\geq$ 9 & 4,226 (0.89\%) \\ \midrule
Total & 476,522 (100.00\%)\\
\bottomrule
\end{tabular}
\end{table}

\begin{figure}
    \centering
    \def\svgwidth{0.8\columnwidth}
    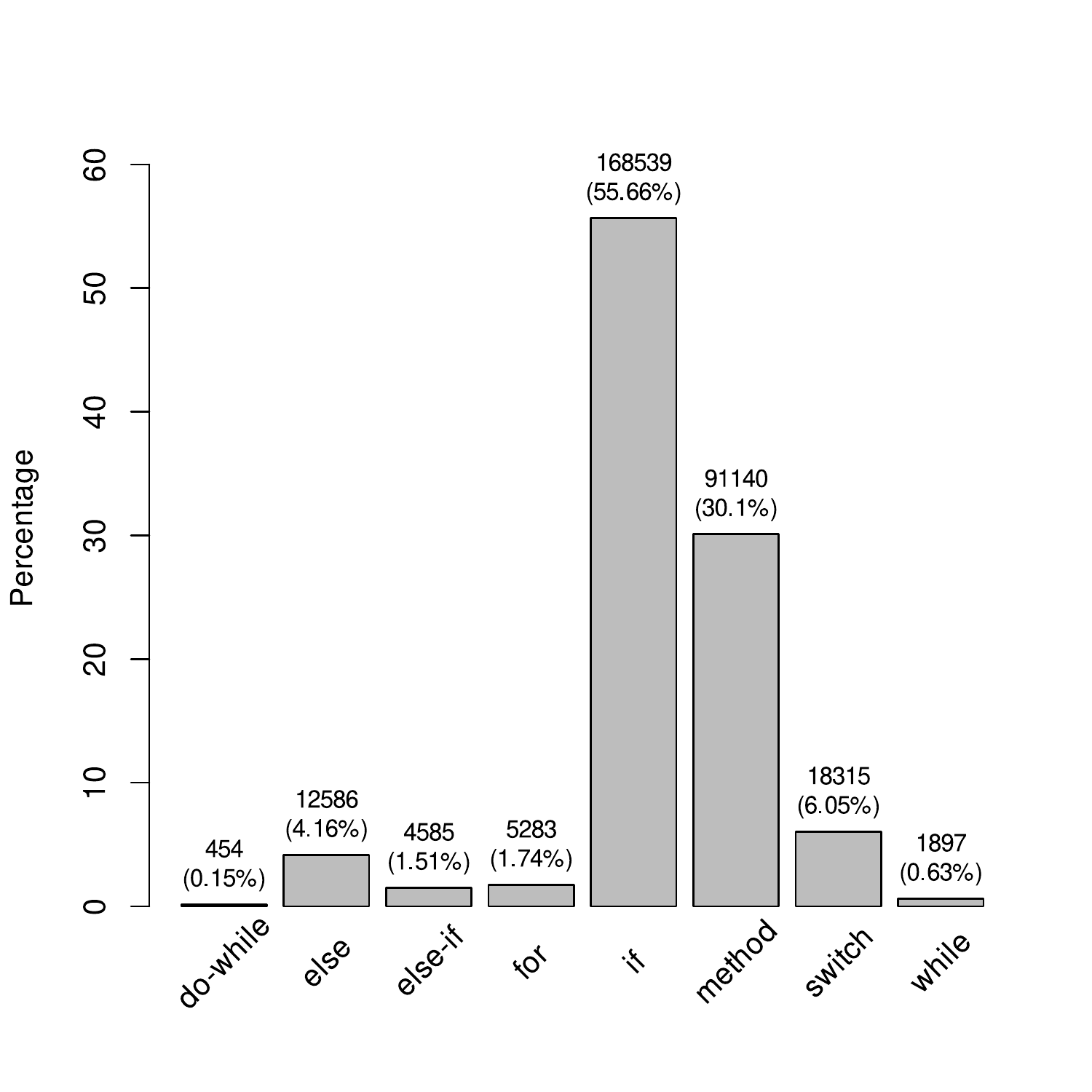
    \caption{Distribution of logging statements across different program constructs.}
    \label{fig:log_dsit_blocks}
\end{figure}

Complementing this information, Figure~\ref{fig:log_dsit_blocks} shows the distribution of logging statements in different program constructs. We found that 55.66\% (168,539 out of 302,799) of logging statements are used inside the \texttt{if} block and  4,585 (1.51\%) are within \texttt{else-if}. Together, they represent 57.17\% of the use of logging. These results are similar to those of the study by~\citet{pecchia2015industry}, who reported that around 60\% of the logging statements are used inside the \texttt{if} blocks. These logs are typically used for logging errors after checking the return value of function calls. The logging statements in the \texttt{else} block accounts for 4.16\% (12,586 out of total 302,799 logging statements). The \texttt{switch} block, which is another control statement available in C, accounts for 6.05\% of the total logging statements. The logging statements used directly inside loop controls represent only 2.52\% of the logging statements. The logging statements used directly within functions (i.e., not in any of the program constructs) account for 30.10\% of the total number of logging statements. \textcolor{black}{This suggests that the cyclomatic complexity of a function may be related to the presence of logging statements. Given that developers may need to understand when a function follows a particular path during the program execution, they add logging to conditional constructs. However, this does not occur within loops. This can be explained by the fact that adding a log statement to a loop may generate a high number of log messages and slow down the function performance.}

\subsubsection{RQ1: Summary}

\paragraph{\textbf{RQ1 - Findings.}} 

The log density in Linux v5.3 is 27, i.e., on average there is one logging statement for every 27 lines of code. This is similar to the study of ~\citet{yuan2012characterizing} on C/C++ systems, but different from the result obtained by ~\citet{chen2017characterizing} when working on Java projects. The logging code represents 3.73\% of the entire source code, with \textit{filesystem} and \textit{drivers} being the most logged subsystems with a log ratio of 4.26\% and 3.97\% and a log density of 23 and 25, respectively. We found that 72.36\% of the total number of files in the Linux kernel have at least one logging statement. However, only 26.12\% of the functions in the Linux kernel have one or more logging statements. Further studies, such as the work from~\cite{li2018studying}, are needed in order to understand which topics of source code are more likely to be logged. 57.17\% of the logging statements are used within an \texttt{if} and \texttt{else-if} blocks, while 30.10\% are used inside code blocks in no particular program constructs.

\paragraph{\textcolor{black}{\textbf{RQ1 - Implications.}}}

\textcolor{black}{The fact that logging code accounts for almost 4\% of the entire code base is a strong reason for developers to  carefully review logging code with the same rigour they use to review the other code. The discrepancies in the log density of various Linux subsystems calls for future studies (with a protocol similar to that by \citet{fu2014developers}) to understand the goals of logging (e.g., debugging,  auditing, security, etc.) in the different parts of the kernel. This study should be extended to the file and function levels where the distribution of logging statements is also heterogeneous with no apparent reason. Another implication of these results is the need for common guidelines for logging to ensure that best logging practices are followed. Finally,  we suggest to study the correlation between code quality (using metrics such as cyclomatic complexity) and the need for logging, which may explain the fact the majority of logging code is found in if-else blocks.}

\subsection{RQ2: Logging code evolution}\label{sec:rq2}

As shown in the previous section, logging is widespread in Linux---for every 27 lines of code, there is one line of logging code. Now, we focus on understanding how the logging code evolves across various releases of the kernel in terms of size as well as the type and the nature of changes.

\subsubsection{Evolution of the size of logging code}\label{log-usage}

\begin{figure}
\begin{subfigure}{.49\textwidth}
    \centering
    \def\svgwidth{\columnwidth}
    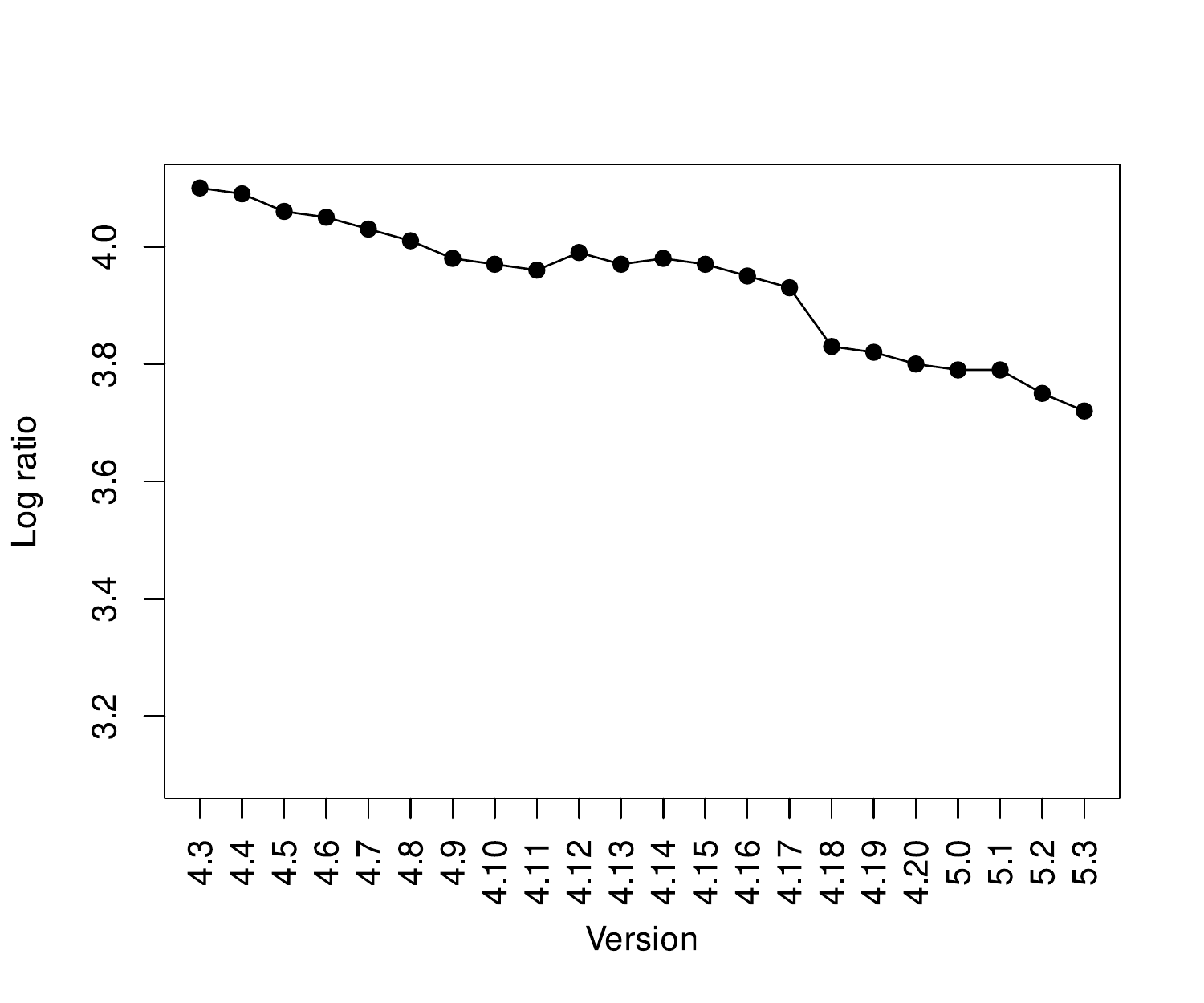
    \caption{Log ratio}
    \label{fig:usage_ind}
\end{subfigure}
\begin{subfigure}{.49\textwidth}
    \centering
    \def\svgwidth{\columnwidth}
    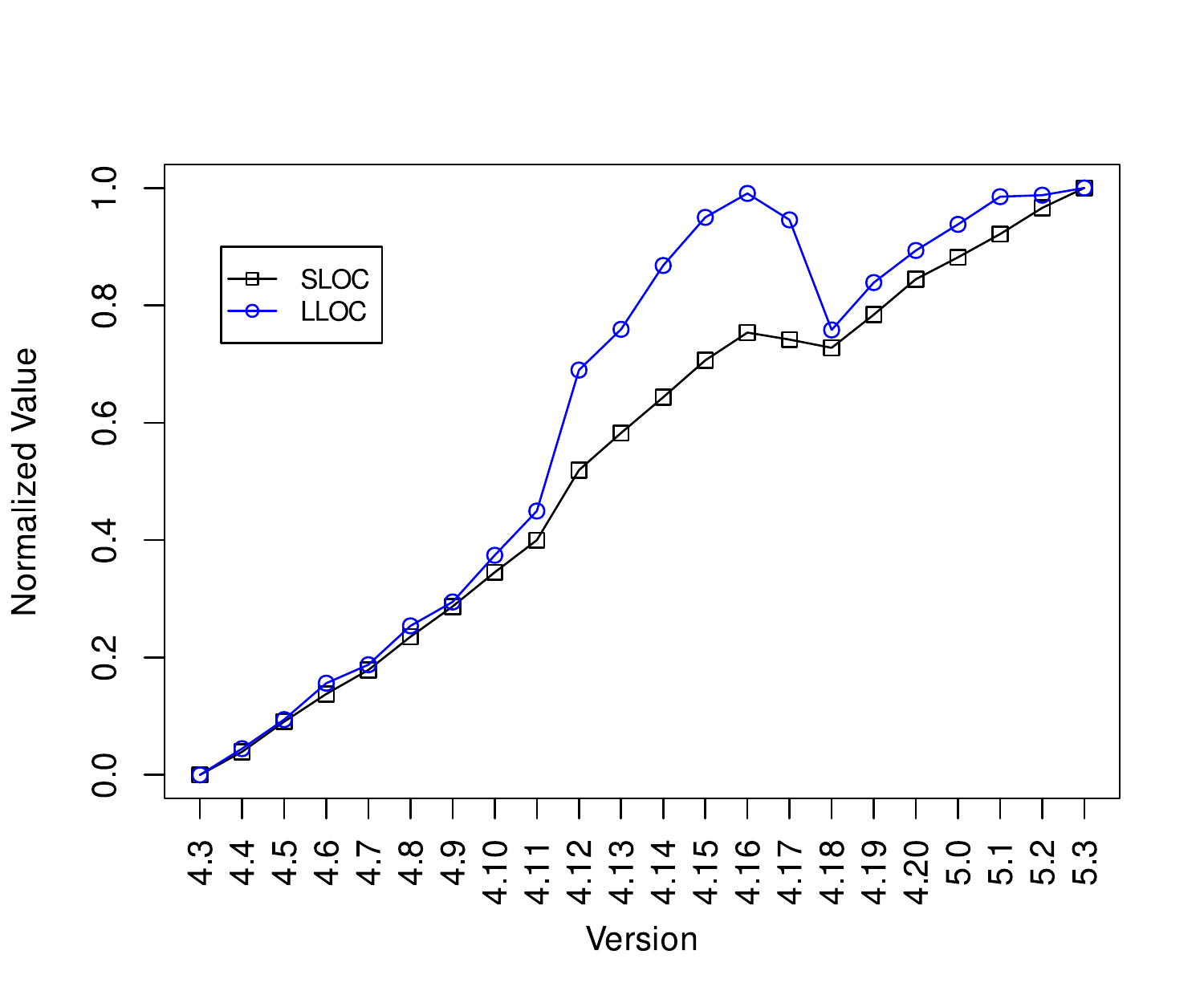
    \caption{Comparison of SLOC and LLOC}
    \label{fig:sloc_vs_lloc}
\end{subfigure}
\caption{Evolution of the use of logging code between Linux v4.3 and v5.3.}
\label{fig:logging_evol}
\end{figure}

We start by analyzing the logging code size by looking at two perspectives. First, we observe how the proportion of logging code evolved, measured by the log ratio metric. Second, we compare the evolution of SLOC and LLOC. The evolution of the log ratio metric for the Linux system from v4.3 to v5.3 can be seen in Figure~\ref{fig:usage_ind}. We observe that it has been decreasing over the years. \textcolor{black}{This can be due to the addition of code that has fewer logging statements than the average or the deletion of code that has more logging statements than the average. To better understand this observed evolution, we complement this analysis with the data shown in} Figure~\ref{fig:sloc_vs_lloc}, which indicates how LLOC and SLOC, individually, increased across the different versions.  SLOC and LLOC are normalized using min-max normalization to fall in the $[0,1]$ interval. 

The SLOC and LLOC curves shown in Figure~\ref{fig:sloc_vs_lloc} generally have a similar behavior, except for versions v4.11 to v4.18. By going through the Linux kernel changelogs, we found that these inconsistencies seem to be the result of the addition/removal of drivers or filesystems, which contained a large amount of logging code. The Linux kernel v4.12 added Intel \emph{atomisp} camera drivers (commit \texttt{a49d253}) and rtl8723bs sdio wifi driver (commit \texttt{554c0a3a}). These two changes increased the number of logging statements by 6,103, which contributed to a sharp increase in the number of lines of logging code. Similarly, we can see a sudden decrease in SLOC and LLOC between v4.16 and v4.18 releases. After inspecting the Linux kernel v4.18 changelogs, we found that SLOC of the Linux kernel v4.18 was smaller than its previous release, and that occurred just three times in the history of the Linux kernel before the release of kernel v4.18. The reason for this can be attributed to removal of the \emph{lustre} filesystem (commit \texttt{be65f9e}) and the \emph{atomisp} driver (commit \texttt{51b8dc51}), which earlier contributed to 10,442 logging statements. \textcolor{black}{As discussed earlier, the amount of logging statements within modules and files varies. Consequently, the removal of particular parts of the system can largely affect the amount of logging code when compared to the rest of the code.}

Considering both the log ratio and the number of lines of logging code, it is thus possible to observe that, though the number of lines of code increased as the Linux kernel evolved, the log ratio decreased. This can be due to a number of reasons, such as the types of changes made in the analyzed Linux versions---as it was discussed, different subsystems have varying amounts of logging. This may also be explained by the reliance on debugging and tracing tools, which can be used to diagnose problems as shown by~\citet{corbet2016tracepoints} and~\citet{jake2019tracepoints}. These authors noticed an increase in the use of tracepoints rather than simple \texttt{printk()} in recent versions of Linux. The term tracing is used here to show the flow of execution of specific program constructs, e.g., traces of routine calls, and system calls \citep{hamou2002compression, hamou2004survey}). A discussion with one Linux expert (see Section 5.2) confirms this finding. We need to conduct further studies including user studies with Linux developers to understand the real causes behind this decline of logging code (see Section 5.3 on the implications for future research for more discussion), \textcolor{black}{including the identification of the impact of the use of new tools that capture runtime information automatically without the need for logging statements added manually. It is relevant to understand what types of statements are removed and what types of statements remain in the code.}

Similarly to previous studies~\citep{yuan2012characterizing,chen2017characterizing,zeng2019studying,li2019guiding}, we calculate the ratio of the number of commits that involve modifications to logging code (i.e., additions, deletions, or updates to logging statements) to the total number of commits so that it is possible to understand how often a source code revision involves a modification to logging code. Out a total of 285,045 commits, we found that 39,351 (14\%) commits involve modifications to logging code. This result compares to that of the study of~\citet{yuan2012characterizing} on C/C++ systems, who found that 18\% of the commits they studied involve log modifications. It is also similar to the results reported by~\citet{chen2017characterizing} who found that there are around 21\% of such commits in the studied Java systems. \textcolor{black}{Some changes are expected because logging statements should be kept consistent with the corresponding code. However, it is relevant to understand why the logging statements change when this is not the case. This is in fact investigated in our RQ3.}

\subsubsection{Types of modifications in logging statements}\label{sec:mod-log-code}

Given that logging code changes overtime, we now look at the types of modifications that are made. Table~\ref{logchangetype} reports the number of logging statements added, deleted, and updated across Linux versions v4.3 to v5.3. There are 211,437 logging statement modifications, out of which 24.78\% are log updates,  45.99\% are log insertions, and log deletions account for 29.23\%. The \textit{drivers} subsystem alone represents 86.60\% of the total modifications made to logging code, followed by \textit{arch} (5.47\%), and \textit{filesystem} (3.74\%). This is somewhat expected because the \textit{drivers} subsystem is considerably larger (over 10 million SLOC) than the other subsystems (see Table~\ref{tbl:rq1:results}). \textcolor{black}{In addition, drivers are typically dynamic in operating systems. As new devices are released or become obsolete, the operating system must be evolved accordingly. Consequently, some of these modifications to the logging statements may be not due to changes specifically done on these statements, but to the addition or deletion of drivers. Moreover, as discussed in our RQ1, drivers typically include much logging code. This justifies why the \textit{drivers} subsystem is associated with the vast majority of modification on logging code.}

\begin{table}
\caption{Distribution of modifications made to logging statements.}
\label{logchangetype}
\centering
\begin{tabular}{lrrr}
\toprule
{\textbf{Subsystem}} &  {\textbf{Insertion}} & {\textbf{Deletion}} & {\textbf{Update}}\\ 
\midrule
arch    & 3,987  &   4,620    & 2,962  \\
core    & 1,863  &   647      & 1,700  \\
driver  & 87,125 &   53,235   & 42,748 \\
fs      & 2,776  &   1,626    & 3,502  \\  
net     & 1,490  &   1,677    & 1,479  \\
\midrule
\textbf{Total}      & 97,241 & 61,805 & 52,391 \\
\textbf{Percentage} & 45.99\% & 29.23\% & 24.78\% \\
\bottomrule
\end{tabular}
\end{table}

The percentage of log additions in Linux is similar to that reported by~\citet{chen2017characterizing}, who showed that log additions contribute to 18\%--41\% of the total log modifications.  \citet{yuan2012characterizing} did not report the percentage of log additions in their examination of C/C++ systems. 

We found that the number of log deletions represents 29.23\% of the total number of modifications made to logging statements. This result differs significantly from the work of~\citet{yuan2012characterizing} who reported that the number of log deletions is only 2\% of the total number of modifications. Our result is similar to that of~\citet{chen2017characterizing} who reported that log deletion contributes to 26\% of the total modifications in Java systems. Such evolution of logging code may adversely impact the bug triaging process as developers rely on the logs contained in the bug reports, as noted by~\citet{anranchen2019icse}. The authors found that it is not possible to rebuild the execution paths for bug reproduction from bug report logs in 34\% cases. They also argued that the continuous evolution of system logs can have an effect on the accuracy of log processing tools and machine learning models deployed for identifying anomalous activities, as models need to be retrained whenever logging statements are changed.

To drill down into the data, we grouped log additions and deletions into two categories: added along with the addition of a new file and deleted along with the deletion of an existing file. Table~\ref{with_file} shows a detailed breakdown. We found that 55.77\% of all log insertions were made along with the addition of new files. Similarly, we found that 55.34\% of the deleted logging statements were deleted along with the deletion of existing files. \textcolor{black}{These data corroborate with our hypothesis of why the \texttt{drivers} subsystem accounts for many modification on logging code. These type of changes---that is, addition and deletion of files---have in fact a large impact in the log ratio, as shown in Figure~\ref{fig:logging_evol}.}


\begin{table}
\caption{Impact of adding or deleting files on logging code.}
\label{with_file}
\centering
\begin{tabular}{lrr}
\toprule
{\textbf{Subsystem}} &  {\textbf{Additions in new files}} & {\textbf{Deletions in a deleted files}}\\
\midrule
arch    &    1,289   & 3,038 \\
core    &    614     & 84    \\
driver  &    51,178  & 30,452 \\
fs      &    833     & 409   \\
net     &    320     & 218   \\
\midrule
\textbf{Total}      & 54,234 & 34,201 \\
 & 55.77\% of the additions & 55.34\% of the deletions \\
\bottomrule
\end{tabular}
\end{table}

\subsubsection{Updates in logging statements}\label{sec:upd-log-code}

We now proceed to a deeper analysis of logging changes, analyzing how log statements are updated. We classify log updates into three categories depending on which part of the logging statement has been modified: (i) the \emph{logging function} (or macro), including the log level, (ii) the \emph{static content} representing the log message, and (iii) the \emph{dynamic content}, i.e.\ the variables and function calls. Table~\ref{updatedistribution} presents the results. Note that one update may consist of one or more changes to the same logging statement. For example, if the log function and the static content of a logging statement has changed, they appear as two updates. This explains why the total number of log updates is higher than the total number of the updated logging statements. We next analyze the different categories of updates. 

\begin{table}
\caption{Distribution of updates made to logging statements.}
\label{updatedistribution}
\centering
\begin{tabular}{lrrr}
\toprule
{\textbf{Subsystem}} &  {\textbf{Logging function}} & {\textbf{Static content}} & {\textbf{Dynamic content}}\\ 
\midrule
arch    &    1,226   &   2,014    &   1,284    \\
core    &    606     &   979 &   721      \\
driver  &    15,685  &   23,127   &   28,501   \\
fs      &    1,344   &   2,422    &   1,863    \\
net     &    435     &   809     &   952      \\
\midrule
\textbf{Total}      &  19,296 & 29,351 & 33,321  \\
\textbf{\% of updates} & 36.83\% & 56.02\% & 63.60\% \\
\bottomrule
\end{tabular}
\end{table}

\paragraph{Changes made to the logging function.} 

Of the 52,391 updated logging statements, 19,296 (36.83\%) include modifications to the used logging functions. A wide range of changes are possible, as shown in Table~\ref{func-change}. In this table, we show only the most frequent changes in logging function, i.e.\ those with a frequency higher than 100.

\begin{table}
  \footnotesize
  \caption{Frequency of changes made to logging functions (frequency $>$ 100).}
  \label{func-change}
  \centering
  \begin{tabular}{llr}
    \toprule
    {\textbf{Old logging function}} & {\textbf{New logging function}}       & {\textbf{Frequency}} \\
    \midrule
    \spverb|printk|        & \spverb|pr_err|              & 950         \\
    \spverb|printk|        & \spverb|pr_info|             & 881         \\
    \spverb|BTC_PRINT|     & \spverb|btc_alg_dbg|         & 663         \\
    \spverb|btc_alg_dbg|   & \spverb|RT_TRACE|            & 621         \\
    \spverb|printk|        & \spverb|pr_cont|             & 606         \\
    \spverb|printk|        & \spverb|pr_warn|             & 591         \\
    \spverb|PDEBUG|        & \spverb|gspca_dbg|           & 477         \\
    \spverb|pr_err|        & \spverb|dev_err|             & 440         \\
    \spverb|pr_info|       & \spverb|ioc_info|            & 418         \\
    \spverb|PDBG|          & \spverb|pr_debug|            & 392         \\
    \spverb|dev_err|       & \spverb|DRM_DEV_ERROR|       & 387         \\
    \spverb|pr_debug|      & \spverb|dev_dbg|             & 369         \\
    \spverb|test_msg|      & \spverb|test_err|            & 345         \\
    \spverb|brcmf_err|     & \spverb|bphy_err|            & 315         \\
    \spverb|RT_TRACE|      & \spverb|pr_err|              & 275         \\
    \spverb|pr_warning|    & \spverb|pr_warn|             & 262         \\
    \spverb|pr_err|        & \spverb|ioc_err|             & 231         \\
    \spverb|printk|        & \spverb|pr_debug|            & 226         \\
    \spverb|BT_ERR|        & \spverb|bt_dev_err|          & 205         \\
    \spverb|DTRACE|        & \spverb|dml_print|           & 173         \\
    \spverb|PRINT_ER|      & \spverb|netdev_err|          & 169         \\
    \spverb|dev_dbg|       & \spverb|musb_dbg|            & 163         \\
    \spverb|pr_info|       & \spverb|dev_info|            & 161         \\
    \spverb|SSI_LOG_DEBUG| & \spverb|dev_dbg|             & 156         \\
    \spverb|BUGMSG|        & \spverb|arc_printk|          & 153         \\
    \spverb|SSI_LOG_ERR|   & \spverb|dev_err|             & 152         \\
    \spverb|dev_info|      & \spverb|dev_dbg|             & 151         \\
    \spverb|dev_err|       & \spverb|dev_dbg|             & 146         \\
    \spverb|DRM_ERROR|     & \spverb|DRM_DEBUG|           & 145         \\
    \spverb|dev_info|      & \spverb|pci_info|            & 139         \\
    \spverb|PERR|          & \spverb|gspca_err|           & 129         \\
    \spverb|btc_iface_dbg| & \spverb|RT_TRACE|            & 119         \\
    \spverb|BTC_PRINT|     & \spverb|btc_iface_dbg|       & 119         \\
    \spverb|pr_info|       & \spverb|pr_debug|            & 119         \\
    \spverb|gvt_err|       & \spverb|gvt_vgpu_err|        & 118         \\
    \spverb|DRM_ERROR|     & \spverb|DRM_DEV_ERROR|       & 104         \\
    \spverb|pr_warn|       & \spverb|dev_warn|            & 102         \\
    \spverb|pr_info|       & \spverb|pr_info_ratelimited| & 102         \\
    \spverb|pr_err|        & \spverb|pr_debug|            & 100         \\
    \bottomrule
  \end{tabular}
\end{table}

The analysis of the changes to the logging functions revealed that 6,512 (33.75\%) of these updates are changes between \spverb|printk|, \spverb|pr_<*>|, and \spverb|dev_<*>| macros. For example, in commit \spverb|26a0a10|, a developer updated logging statements from \spverb|printk| to using device-aware \spverb|dev_err()/dev_info()| logging functions for improving the precision of the resulting logs by including device-specific information. This is further cemented by an observation that there has been a steady decrease in the usage of the \texttt{printk()} function, with a usage reduction of 29.32\% between versions 4.3 and 5.3 of the Linux kernel. However, we found that the use of \texttt{pr\_*()} and \texttt{dev\_*()} functions has increased to replace \texttt{printk()} call sites~\citep{corbet2012perils}. \textcolor{black}{This gives evidence of evolving the importance of logging code in Linux, that is, developers have been modifying logging functions, possibly with the aim of creating standardised functions.}

We also found an increasing use of the so-called ``rate limited'' logging functions such as the \spverb|<*>[_once/_ratelimited]| family of macros, which can be seen in commit \spverb|527aa2a|, where all calls to \spverb|pr_info| were converted to \spverb|pr_info_ratelimited|. The objective of this type of functions is to prevent overloading the log buffers by controlling the amount of logs generated in a given period of time. \textcolor{black}{Consequently, while writing logging code, it is important to consider not only the information being logged, but also the amount of information being generated and its impact on the application execution.}

Moreover, we observed that many changes to logging functions are triggered by the need to make them more concise. For example, in commit \spverb|466414a|, a developer introduced \spverb|btc_alg_dbg| and \spverb|btc_iface_dbg| logging macros, and converted all calls to \spverb|BTC_PRINTK| to the new functions (\spverb|btc_<*>_dbg|). The benefit  is that software developers do not have to specify \spverb|btc_msg_type|, resulting in more concise logging statements. Ten months later, in commit \spverb|10468c3|, all calls to \spverb|btc_<*>_dbg| were again changed to another logging function, named \spverb|RT_TRACE|, to be consistent with the use of this function in other drivers. 

We thus conclude from the above observations that changes to logging functions aimed to improve the quality of the logging output by either enhancing precision, conciseness, or consistency. However, after analyzing many commits related to log updates, we could not find any evidence that there was a Linux-wide strategy, which suggests that these updates are a result of established guidelines. It appears that the decision on how to log is left to the discretion of the developers. 

The second reason for logging function modification is changing the severity of logging statements by specifying log levels. The \spverb|printk| function allows one of the eight log levels defined in \path{/include/linux/kern_levels.h}. For example, 
\begin{verbatim}
  printk(KERN_ERR "GCT Node MAGIC incorrect - GCT invalid\n");
\end{verbatim}
However, the new logging API introduced in Linux 1.3.98\footnote{\url{https://repo.or.cz/davej-history.git?a=commit;h=aa66269c}} embedded log levels into the function names, such as \spverb|pr_debug| and \spverb|pr_info|. \textcolor{black}{This change makes the logging API richer, as developers have dedicated functions for specific log levels.}

Table~\ref{verbchnage} provides the distribution of changes made to the severity of logging statements. If the developer updates a logging statement to use a new logging function, while keeping the same log level, we do not consider this change as a log level change. \textcolor{black}{This avoids including in the analysis refactoring in the logging code that preserved the log level.} As stated in previous work~\citep{yuan2012characterizing}, developers often fail at determining how critical an error is in the first attempt. This observation is confirmed by the finding from our study that, out a  total of 4,127 log level changes, approximately one third of the total log level changes were between ERR and DEBUG log levels. Specifically, 720 (17.45\%) logging statements lowered the severity of the log message from ERR to DEBUG, and 616 (14.93\%) from DEBUG to ERR. We found a total of 1,522 (36.88\%) instances where developers increased the severity of a logging statement to increase their visibility. 
In addition, logging debugging messages at the ERROR level would result in log flooding, making it difficult to diagnose the real problems. 
We found  that a total of 2,605 (63.12\%) instances where  developers reduced the severity of a logging statement in order to prevent log flooding. In fact, of these 4,127 log level modifications, 3,832 (92.85\%) changes are between ERR, WARNING, INFO, and DEBUG log levels. \textcolor{black}{This suggests, as stated before, a lack of standards when logging, in particular, when choosing a log level. Moreover, it indicates that it is by checking generated log messages generated by already written logging statements that developers are better able to reason about appropriate log levels.}

\begin{table}
  \scriptsize
  \caption{Changes in the level of logging statements between Linux v4.3 and v5.3.}
  \label{verbchnage}
  \centering
  \begin{tabular}{lrrrrrrrr}
    \toprule
    {\diagbox[width=6em]{Old}{New}} & {EMERG}    & {ALERT}    & {CRIT}     & {ERR}      & {WARN}     & {NOTICE}   & {INFO}     & {DEBUG}    \\
    \midrule
    EMERG                           & \textbf{0} & 1          & 10         & 51         & 30         & 0          & 1          & 0          \\
    ALERT                           & 0          & \textbf{0} & 6          & 8          & 10         & 0          & 0          & 0          \\
    CRIT                            & 1          & 0          & \textbf{0} & 8          & 3          & 0          & 0          & 0          \\
    ERR                           & 1          & 0          & 5          & \textbf{0} & 242        & 19         & 195        & 720        \\
    WARN                         & 22         & 2          & 14         & 343        & \textbf{0} & 14         & 336        & 346        \\
    NOTICE                          & 0          & 0          & 0          & 10         & 5          & \textbf{0} & 21         & 22         \\
    INFO                            & 1          & 0          & 1          & 188        & 62         & 23         & \textbf{0} & 562        \\
    DEBUG                           & 0          & 0          & 0          & 616        & 77         & 6          & 145        & \textbf{0} \\
    \bottomrule
  \end{tabular}
\end{table}

\paragraph{Changes made to the static content.} 

56.02\% of the 52,391 changed logging statements, \textcolor{black}{i.e.\ the majority}, include modifications to the static content (i.e.\ the log message).  A similar result has also been observed by~\citet{chen2017characterizing} and~\citet{yuan2012characterizing}, who reported a ratio of 14\%--65\% and 18\%--56\%, respectively. Prior studies list \textit{fixing inconsistency}, \textit{clarification}, and \textit{spelling/grammar mistakes} as the major causes of this type of modifications~\citep{chen2017characterizing,yuan2012characterizing,chen2019extracting}. \textcolor{black}{Because static content consists of qualitative data, we are not able to make further claims about these changes in RQ2. However, this is further investigated in RQ3, with a qualitative analysis of logging modifications.}

\paragraph{Changes made to the dynamic content.} 

Of the 52,391 updated logging statements, 33,321 (63.60\%) include modifications to the dynamic content (i.e.\ variables and function calls). \citet{yuan2012characterizing} found that developers often add variables into existing logging statements as afterthoughts, which can aid in the failure diagnosis process. However, changes to dynamic content represent only 27\% of all log modifications in their study. Our finding is, however, in line with the study of~\citet{li2019guiding} on 12 C/C++ open-source projects, where the authors found that 69.1\% updates to logging statement made modifications to the dynamic content. This high number of changes to dynamic content has also been observed by~\citet{chen2017characterizing} and~\citet{zeng2019studying}, in which they studied server/desktop and Android applications written in Java, respectively. \textcolor{black}{That is, despite \citet{yuan2012characterizing} already claims that the addition of variables to logging statement is frequent, in various projects---including Linux---it is more than twice more frequent than they observed in their target systems.}

One possible explanation for this high number of changes of the logging statement dynamic content in Linux can be seen in commit \spverb|6be9005|, where the developer switched to \spverb|DRM_DEV_DEBUG_<*>| instead of \spverb|DRM_DEBUG|. \spverb|DRM_DEV_DEBUG_<*>| are device-aware logging macros and they require \spverb|struct *device| as an argument to include device name in the log output. In order to help developers decide which variables should be included in the logging statement, a recent study by~\cite{varstolog2019} proposed a deep learning-based approach that achieved an average MAP score of 0.84 on nine open-source Java projects. Such approaches tailored to Linux kernel domain could be helpful to alleviate this problem of what information should be included in a logging statement.

\subsubsection{RQ2: Summary}

\paragraph{\textbf{RQ2 - Findings.}} 

The log ratio (LLOC/SLOC) has been decreasing slowly from v4.3 to v5.3, with a log ratio of 4.10\% and 3.72\%, respectively. This represents a reduction of 9.27\%. SLOC and LLOC are closely correlated across most versions of Linux kernel between v4.3 and v5.3. This seems to be caused by the increase use of tracing mechanism as a substitute to logging. We found that 14\% of the commits made between the Linux kernel versions v4.3 and v5.3 involve modifications to logging code. Out of the 211,437 logging statements modifications, 24.78\% are log updates, 45.99\% are log insertions, and 29.23\% are log deletions. The majority of changes to logging code (86.60\%) are made in the \textit{drivers} subsystem. The changes to logging functions are triggered by the need to improve the quality of the logging output by either enhancing precision, conciseness, or consistency.
We found that 92.85\% of the changes of log levels are between ERR, WARNING, INFO, and DEBUG log levels, suggesting that it is difficult for Linux developers to decide on which log level to use. We found that 63.60\% of the updated logging statements include modifications to the dynamic content, while 56.02\% are changes to the static content.

\paragraph{\textcolor{black}{\textbf{RQ2 - Implications.}}}

\textcolor{black}{A study should be conducted to dig deeper into the reasons behind the decline of the log ratio over the years. The study should explore the correlation between logging and tracing as it appears that tracing is being used as a substitute to logging in more recent versions of Linux. We recommend to examine the benefits of logging, tracing, and other debugging mechanisms when used individually and together. This may lead to better ways to optimize the use of these mechanisms. For example, tracing has the advantage of being less dependent on the developer's input, and hence requires less maintenance. However, it comes with an added overhead due to the presence of trace points, which may limit its overall deployment. Another implication of these results is related to the lack of  guidelines and  best practices, which is the main trigger behind the many changes to logging statements. This is also evidenced in the high number of changes to log levels, which clearly show that developers face challenges when deciding on the log levels to use. These issues should be addressed by developing a set of guidelines and best practices that do not only consider the information being logged (i.e., what to log), but also the amount of information that is generated and its impact on the application execution. Finally, the high number of changes to the dynamic variables is problematic since many errors can occur, leading to faults and crashes. To address this issue, we suggest to use tools that would automatically recommend updates to the dynamic variables of logging statements when these variables have been modified in others parts of the code.}

\subsection{RQ3: Afterthought changes in logging code}\label{sec:rq4}

For this research question, we look into \emph{afterthought changes}, which are those that explicitly address a bug caused by or is used to enhance the logging code~\citep{yuan2012characterizing,chen2017characterizing}. It is essential to study afterthought changes because they add to the overall maintenance effort. Having a large number of afterthought log changes may defeat the very purpose of logging, which is to reduce the maintenance effort by facilitating debugging and other failure diagnosis tasks. 

An example of a commit addressing a problem associated with logging code can be seen in Listing~\ref{lst:d618651}. In this code snippet, there is an error message displayed by the \emph{thinkpad\_acpi} driver when brightness interfaces are not supported, encouraging the user to contact IBM for this problem. However, according to the developer that handled this commit, back-light interfaces on newer devices are supported by the \emph{i915} driver. The developer decided to change the log level from ``\emph{error}'' to ``\emph{info}'' to reduce the visibility of the log message. Another example consists of a commit in which a developer decided to enhance the existing logging statements by including \spverb|qp_num(qp)| in debug messages to improve debugging tasks (commit \spverb|e404f94|).

\noindent\begin{minipage}[t]{\textwidth}
\begin{lstlisting}[language=C, caption={Commit \texttt{d618651} - thinkpad\_acpi: Don't yell on unsupported brightness interfaces.}, label=lst:d618651, captionpos=b, breaklines=true,
  postbreak=\mbox{\textcolor{gray}{$\hookrightarrow$}\space}]
--- a/drivers/platform/x86/thinkpad_acpi.c
+++ b/drivers/platform/x86/thinkpad_acpi.c
@@ -6459,8 +6459,7 @@ static void __init tpacpi_detect_brightness_capabilities(void)
        pr_info("detected a 8-level brightness capable ThinkPad\n");
        break;
    default:
-       pr_err("Unsupported brightness interface, "
-           "please contact %s\n", TPACPI_MAIL);
+       pr_info("Unsupported brightness interface\n");
        tp_features.bright_unkfw = 1;
        bright_maxlvl = b - 1;
    }
\end{lstlisting}
\end{minipage}

To understand the nature of afterthought changes made to logging code, we conducted a qualitative analysis, detailed in Section~\ref{sec:study-design}, in which we manually examined the corresponding fixes provided by the Linux developers. The ultimate goal is to gain deep insight into the reasons behind these changes, which can help developers and researchers design new approaches and tools to prevent these problems. To analyze the resulting 900 commits, for each commit, the first author reviewed the commit message, the commit diff, related artefacts such as bug reports, and discussions on the Linux kernel mailing lists, if available. He classified the reasons behind these afterthought changes into 13 categories, which were reviewed and validated by the other authors. These categories are shown in Table~\ref{rq4table} and are discussed in more detail in the subsequent sections, split into five groups.

\begin{table}
    \caption{Characterization of fixes to the logging code.}
    \label{rq4table}
    \centering
    \begin{tabular}{l l r}
        \toprule
        \textbf{Group} & \textbf{Logging Fix} & \textbf{\#Commits} \\ 
        \midrule
        Lack of Information	& LF01: Missing variables and statements & 92 \\
        					& LF02: Imprecise logging messages & 62 \\
		\hline
        Issues in 			& LF03: Null pointer dereference & 13 \\
        variable usage		& LF04: Uninitialized variables & 9 \\
        					& LF05: Logging wrong variables & 28 \\
		\hline
        Inadequate logging	& LF06: Fixing incorrect log levels & 156 \\
        configuration		& LF07: Fixing inconsistencies & 163 \\
        					& LF08: Deleting redundant information & 31\\ 
		\hline
		Writing issues		& LF09: Language mistakes & 171 \\
        					& LF10: Fixing format specifiers & 125\\ 
        					& LF11: Message mistakes & 6 \\
        					& LF12: Formatting issues & 71 \\ 
		\hline
        Disclosure of & LF13: Revealing kernel pointers & 16 \\
        Sensitive data	& \\
        \midrule
        \textbf{Total} & & \textbf{900} \\ 
        \bottomrule
    \end{tabular}
\end{table}

\subsubsection{Lack of information}

\begin{table}
    \caption{Examples of lack of information.\label{tbl:rq3LackOfInfo}}
    \setlength{\fboxsep}{0.1pt}
    \centering
    \begin{tabularx}{\textwidth}{lX}
    \toprule
        Example 1 & \textbf{tools/libbpf: improve the pr\_debug statements to contain section numbers} $\vert$ linux\textbf{@}077c066\\ 
        \midrule
        Original & \begin{lstlisting}[style=tblstyle]^^J
        pr_warning("failed to alloc name for prog under section \%s\\n", section_name);^^J
        \end{lstlisting} \\
        Updated & \begin{lstlisting}[style=tblstyle]^^J
        pr_warning("failed to alloc name for prog under section $\colorbox{mygray}{(\%d)}$ \%s\\n", $\colorbox{mygray}{idx}$, section_name);^^J
        \end{lstlisting} \\
        \midrule
        Example 2 & \textbf{ASoC: tas6424: Print full register name in error message} $\vert$ linux\textbf{@}919869214 \\ 
        \midrule
        Original & \begin{lstlisting}[style=tblstyle, mathescape=true]^^J
        dev_err(dev, "failed to read\ $\colorbox{mygray}{FAULT1}$ register: \%d\\n", ret);^^J
        \end{lstlisting} \\ 
		Updated & \begin{lstlisting}[style=tblstyle]^^J
		dev_err(dev, "failed to read\ $\colorbox{mygray}{GLOB\_FAULT1}$ register: \%d\\n", ret);^^J
		\end{lstlisting} \\
        \midrule
        Example 3 & \textbf{mmc: dw\_mmc: fix misleading error print if failing to do DMA transfer} $\vert$ linux\textbf{@}d12d0cb \\ 
        \midrule
        Original & \begin{lstlisting}[style=tblstyle]^^J
        /* We can't do DMA */^^J
		$\colorbox{mygray}{dev\_err}$(host->dev,\ $\colorbox{mygray}{"\%s: failed to start DMA.\textbackslash n"}$, __func__);^^J
		\end{lstlisting} \\
		Updated	& \begin{lstlisting}[style=tblstyle]^^J
		/* We can't do DMA, try PIO for this one */^^J
		$\colorbox{mygray}{dev\_dbg}$(host->dev,\ $\colorbox{mygray}{"\%s: fall back to PIO mode for current transfer\textbackslash n"}$, __func__);^^J
		\end{lstlisting} \\	
		\midrule
		Example 4 & \textbf{x86/fault: Reword initial BUG message for unhandled page faults} $\vert$ linux\textbf{@}f28b11a \\
		\midrule
		Original & \begin{lstlisting}[style=tblstyle]^^J
		pr_alert($\colorbox{mygray}{"BUG: unable to handle kernel \%s at \%px\textbackslash n"}$, address < PAGE_SIZE ? "NULL pointer dereference" : "paging request", (void*)address);^^J
		\end{lstlisting} \\
		Updated	& \begin{lstlisting}[style=tblstyle]^^J
		if (address < PAGE_SIZE \&\& !user_mode(regs))^^J
		\ \ pr_alert($\colorbox{mygray}{"BUG: kernel NULL pointer dereference,}$\ $\colorbox{mygray}{address = \%px\textbackslash n"}$, (void *)address);^^J
		else^^J
		\ \ pr_alert("BUG: unable to handle page fault for address = \%px\\n", (void *)address);^^J
    	\end{lstlisting} \\
        \bottomrule
    \end{tabularx}
\end{table}

\paragraph{LF01: Missing variables and statements.}

We found that 10.22\% of the studied commits are about improving debuggability by adding  information to logging code with the aim of reducing the time needed to diagnose  program failures. This is in line with the finding of~\cite{yuan2012characterizing}, who concluded that developers often add information to the existing logging statements to narrow down the root causes of the underlying problems. For example, in commit \spverb|d0de579|, a developer mentioned: ``\emph{Identify Namespace failures are logged as a warning but there is not an indication of the cause for the failure. Update the log message to include the error status.}'' Another example is commit \spverb|077c066|, which added to the logging statement a local variable representing a section index (\spverb|idx|), as shown in Example 1 of Table~\ref{tbl:rq3LackOfInfo}. The reason for this change is given by the developer as: ``\emph{While debugging a bpf ELF loading issue, I needed to correlate the ELF section number with the failed relocation section reference. Thus, add section numbers/index to the pr\_debug.}'' We also found many cases where developers added logging statements to record additional runtime information. This is exemplified by commit \spverb|9ef8690|, where a developer inserted few additional logging points in order to make errors more visible: ``\textit{The NCSI driver is mostly silent which becomes a headache when trying to determine what has occurred on the NCSI connection. This adds additional logging in a few key areas such as ...}''

\paragraph{LF02: Imprecise logging messages.} 

Ambiguity in an error message can delay the process of diagnosis as it does not allow end users to easily uncover the part of the programs that failed as in, for instance, Example 2 of Table~\ref{tbl:rq3LackOfInfo}. In this example, a developer chose to use the full register name in the error message because a short version of the register name may be ambiguous when diagnosing a fault. A similar example can be found in commit \spverb|2e5d04dad|, where the \textit{iwlagn} driver uses exactly the same error message in three different functions. Therefore, the developer chose to add the name of the function to the error string to disambiguate from where the error originated.

Another common problem is the ambiguity of a log message, which can be  misleading during the analysis phase. Example 3 of Table~\ref{tbl:rq3LackOfInfo} shows an example where the original log message ``\spverb|failed to start DMA|'' together with an error log level might mislead users to think that a fatal error occurred. In fact, although DMA could not be used, the transfer could still be completed by PIO mode. The updated message thus makes this clear as well as has a debug log level.

Another example of this type of logging fix is when developers reword the logging message to make the logs more informative and facilitate analysis. This is shown in Example 4 of Table~\ref{tbl:rq3LackOfInfo}, where developers decided to reword NULL pointer dereference message. The logging message was modified to drop ``\emph{unable to handle}'' from the message, because it might imply that in some cases the kernel actually handles NULL pointer dereference, which is not valid. A similar example is found in commit \spverb|135e535|, where the developer clarified an error message to avoid user confusion. He reported that: ``\emph{Some user who install SIGBUS handler that does longjmp out therefore keeping the process alive is confused by the error message \lstinline[language=C, breaklines=true,basicstyle={\normalsize\tt}]![188988.765862] Memory failure: 0x1840200: Killing cellsrv:33395 due to hardware memory corruption!. Slightly modify the error message to improve clarity.}'' In conclusion, poorly worded log messages may lead to user confusion and fixing these logs takes up maintenance time and effort.

Using the same static text in a single file is also a case that contributes to ambiguity in logging statements. This practice of duplication in logging code was rightfully reported by~\cite{li2019dlfinder} as a logging \textit{code smell}. One example of this practice is depicted in commit \spverb|a15e824| and commit \spverb|90cc7f1|, where the developer added additional information so that log messages can be uniquely identified using search techniques.

\subsubsection{Issues in variable usage}

\begin{table}
    \caption{Examples of issues in variable usage.\label{tbl:rq3Vars}}
    \setlength{\fboxsep}{0.1pt}
    \centering
    \begin{tabularx}{\textwidth}{lX}
    \toprule
        Example 5 & \textbf{staging: ks7010: don't print \lstinline[language=C, basicstyle={\ttfamily}]|skb->dev->name| if skb is null} $\vert$ linux\textbf{@}95d2a32 \\ 
            \midrule
        Original & \begin{lstlisting}[style=tblstyle]^^J
        printk(KERN_WARNING "$\colorbox{mygray}{\%s}$: Memory squeeze, dropping packet.\\n", $\colorbox{mygray}{skb->dev->name}$);^^J
        \end{lstlisting} \\
        Updated & \begin{lstlisting}[style=tblstyle]^^J
        printk(KERN_WARNING "$\colorbox{mygray}{ks\_wlan}$: Memory squeeze, dropping packet.\\n");^^J
        \end{lstlisting} \\
        \midrule
        Example 6 & \textbf{staging: fsl-dpaa2/eth: Don't use netdev\_err too early} $\vert$ linux\textbf{@}0f4c295 \\ 
        \midrule
        Original & \begin{lstlisting}[style=tblstyle]^^J
        $\colorbox{mygray}{netdev\_err}$($\colorbox{mygray}{net\_dev}$, "Failed to configure hashing\\n");^^J
        \end{lstlisting} \\
    	Updated & \begin{lstlisting}[style=tblstyle]^^J
    	$\colorbox{mygray}{dev\_err}$($\colorbox{mygray}{dev}$, "Failed to configure hashing\\n");^^J
    	\end{lstlisting} \\
        \midrule
    	Example 7 & \textbf{btrfs: tree-log.c: Wrong printk information about namelen} $\vert$ linux\textbf{@}286b92f\\ 
        \midrule
        Original & \begin{lstlisting}[style=tblstyle]^^J
        btrfs_crit(fs_info, "invalid dir item name len: \%u", (unsigned)$\colorbox{mygray}{btrfs\_dir\_data\_len}$(leaf, dir_item));^^J
        \end{lstlisting} \\
    	Updated & \begin{lstlisting}[style=tblstyle]^^J
    	btrfs_crit(fs_info, "invalid dir item name len: \%u", (unsigned)$\colorbox{mygray}{btrfs\_dir\_name\_len}$(leaf, dir_item));^^J
    	\end{lstlisting} \\
        \bottomrule
    \end{tabularx}
\end{table}

\paragraph{LF03: Null pointer dereference.} 

In 1.44\% of the cases, a developer attempted to dereference a pointer that may have a NULL value or an empty variable. This is something that can occur in Example 5 of Table~\ref{tbl:rq3Vars}, which illustrates a fix made in commit \spverb|95d2a32|.\footnote{Note that the change made to the logging statement is not semantically equivalent to the original statement, but with the bug fixed. The goal in the change was to fix the NULL pointer and the change was accepted to fix this issue.} In this commit, the original logging statement included the dereferencing of the \lstinline[language=C, basicstyle={\tt}]|skb| pointer, which can be NULL. A NULL pointer dereferencing causes a runtime crash. To prevent this type of problems, developers should incorporate tools such as Coccinelle\footnote{\url{http://coccinelle.lip6.fr/rules/\#null}} in their workflow for detecting dereferences of NULL pointers.

\paragraph{LF04: Uninitialized variables.} 

Out of 900 commits, we found 9 cases in which log messages refer to device names before they are registered. These logs contain messages referring to ``\texttt{unnamed net device}'' or ``\texttt{uninitialized}'', which lead to logs that may confuse end users. Example 6 of Table~\ref{tbl:rq3Vars} shows a fix to one of these problems. The logging statements consists now of a call to the function \texttt{dev\_err}, which considers the possibility that the first parameter is null. In this case, it will add ``\texttt{NULL net\_device}'' to the log message.

\paragraph{LF05: Logging wrong variables.}

Developers specified the wrong variables as arguments of the logging function calls in 28 cases. The root causes of these issues are simple copy/paste mistakes or typographical errors.\footnote{\url{https://cwe.mitre.org/data/definitions/688.html}} This is illustrated in Example 7 of Table~\ref{tbl:rq3Vars}, in which the original and updated logging functions have similar names, but the former calls the function \spverb|btrfs_dir_data_len|, reporting \spverb|data_len| in the error message. However, the correct function is \spverb|btrfs_dir_name_len|, which reports \spverb|name_len| in the message. Such errors are difficult to detect using static analysis tools, because both \spverb|btrfs_dir_name_len| and \spverb|btrfs_dir_data_len| have the same return type.

\subsubsection{Inadequate logging configuration}

\begin{table}
    \caption{Examples of inadequate logging configuration.\label{tbl:rq3config}}
    \setlength{\fboxsep}{0.1pt}
    \centering
    \begin{tabularx}{\textwidth}{lX}
    \toprule
        Example 8 & \textbf{i2c: imx: notify about real errors on dma \lstinline[language=C, basicstyle={\ttfamily}]|i2c_imx_dma_request|} $\vert$ linux\textbf{@}5b3a23a \\ 
        \midrule
        Original & \begin{lstlisting}[style=tblstyle]^^J
        $\colorbox{mygray}{dev\_dbg}$(dev, "can't configure rx channel\\n");^^J
        \end{lstlisting} \\
        Updated & \begin{lstlisting}[style=tblstyle]^^J
        $\colorbox{mygray}{dev\_err}$(dev, "can't configure rx channel (\%d)\\n", ret);^^J
        \end{lstlisting} \\
        \midrule
        Example 9 & \textbf{mei: bus: remove redundant uuid string in debug messages} $\vert$ linux\textbf{@}2bcfdc2 \\ 
        \midrule
        Original & \begin{lstlisting}[style=tblstyle]^^J
        dev_dbg(\&cldev->dev, "running hook \%s\ $\colorbox{mygray}{on \%pUl}$\\n", __func__,\ $\colorbox{mygray}{mei\_me\_cl\_uuid(cldev->me\_cl)}$);^^J
        \end{lstlisting} \\
        Updated & \begin{lstlisting}[style=tblstyle]^^J
        dev_dbg(\&cldev->dev, "running hook \%s\\n", __func__);^^J
        \end{lstlisting} \\
        \bottomrule
    \end{tabularx}
\end{table}

\paragraph{LF06: Fixing incorrect log levels.}

We found that in 17.33\% of the cases there are changes in the log levels. This includes cases where developers failed to make a distinction between fatal errors and errors that are recoverable. Moreover, using a non-error log level for logging error conditions would make it hard to diagnose such errors as the corresponding error messages would go unnoticed. Likewise, logging debugging messages as errors will also result in a flood of log messages making it hard to concentrate on the real problems. In example of such an issue is shown in Example 8 (Table~\ref{tbl:rq3config}), where the developer mentioned that ``\emph{... In contrast real problems that were only emitted at debug level before should be described at a higher level to be better visible and so understandable.}'' Thus, the developer decided to change the log level from \texttt{DEBUG} to \texttt{ERROR}. 

Developers need to carry out logging activities taking into account performance constraints~\citep{ding2015log2,sigelman2010dapper}. One of the most frequent issues is log spamming, which often leads to degradation of system performance. This occurred in commit \spverb|7f20d83|, where the developer downgraded the log message to \texttt{DEBUG} level to suppress frequent ``\emph{VTU miss violations}'' messages, as ``\emph{VTU miss violations}'' are rather common. 

\paragraph{LF07: Fixing inconsistencies.}

We found 163 commits (18.11\%) that were the result of ongoing modernization of logging code. Usually, such improvements do not address severe problems associated with the logging code; instead, they arise from attempts to improve the consistency of logging code across various parts of the Linux kernel. One such change is the use of device-managed logging macros in order to simplify error handling, reduce source code size, improve readability, and/or reduce the risks of bugs.
Another improvement is the use of \texttt{\_\_func\_\_} and \texttt{pr\_fmt} rather than hard-coded function names and module names in error messages. For example, in commit \spverb|a8ab042|, a developer mentioned: ``\emph{Instead of having the function name hard-coded (it might change and we forgot to update them in the debug output) we can use} \texttt{\_\_func\_\_} \emph{instead and also shorter the line so we do not need to break it.}'' As~\cite{yuan2012characterizing} pointed out, inconsistency in the function names referred in the log message is one of the main reasons for changes made to logging code.

\paragraph{LF08: Deleting redundant information.}

We found 31 cases in which developers report information that is redundant or not needed. One illustration of this case found in commit \spverb|2bcfdc2| is shown in Example 9 (Table~\ref{tbl:rq3config}), where the developer removed \emph{uuid} from the debug messages in \path{bus-fixup.c} as this was already part of the device name. Another common pattern that we observed is the removal of \spverb|__func__| from \spverb|dev_dbg()| calls. The reason for this change is given in commit \spverb|b814735| as ``\emph{Dynamic debug can be instructed to add the function name to the debug output using the +f switch, so there is no need for the nfit module to do it again. If a user decides to add the +f switch for nfit's dynamic debug this results in double prints of the function name ..... Thus remove the stray \texttt{\_\_func\_\_} printing.}'' In addition,  removing \spverb|__func__| from \texttt{dev\_*()} callsites helps reduce the Linux kernel size as  pointed out by Wolfram Sang, the current maintainer of the Linux I2C subsystem.\footnote{\url{https://git.kernel.org/pub/scm/linux/kernel/git/wsa/linux.git/commit/?h=strings/rtc-no-func&id=762c5af234c5b816b7da3687a3e703cf8cdc2214}}

\subsubsection{Writing issues}

\begin{table}
    \caption{Examples of writing issues.\label{tbl:rq3writing}}
    \setlength{\fboxsep}{0.1pt}
    \centering
    \begin{tabularx}{\textwidth}{lX}
    \toprule
        Example 10 & \textbf{usbip: vhci: fix spelling mistake: ``synchronuously'' $\rightarrow$ ``synchronously''} $\vert$ linux\textbf{@}cb48326 \\ 
        \midrule
        Original & \begin{lstlisting}[style=tblstyle]^^J
        dev_dbg(\&urb->dev->dev, "urb seq# \%u was unlinked \%s$\colorbox{mygray}{synchronuously}$\\n",seqnum,status == -ENOENT ? "":"a");^^J
        \end{lstlisting} \\
        Updated & \begin{lstlisting}[style=tblstyle]^^J
        dev_dbg(\&urb->dev->dev, "urb seq# \%u was unlinked \%s$\colorbox{mygray}{synchronously}$\\n",seqnum,status == -ENOENT ? "":"a");^^J
        \end{lstlisting} \\
        \midrule
        Example 11 & \textbf{ti\_sci: Use \%pS printk format for direct addresses} $\vert$ linux\textbf{@}595f3a9 \\ 
        \midrule
        Original & \begin{lstlisting}[style=tblstyle]^^J
        dev_err(dev, "Mbox timedout in resp(caller:\ $\colorbox{mygray}{\%pF}$)\\n", (void *)_RET_IP_);^^J
        \end{lstlisting} \\
        Updated & \begin{lstlisting}[style=tblstyle]^^J
        dev_err(dev, "Mbox timedout in resp(caller:\ $\colorbox{mygray}{\%pS}$)\\n", (void *)_RET_IP_);^^J
        \end{lstlisting} \\
        \midrule
        Example 12 & \textbf{greybus: gpb: Fix print mistakes} $\vert$ linux\textbf{@}b908dec \\ 
        \midrule
        Original	& \begin{lstlisting}[style=tblstyle]^^J
        if (gb_usb_protocol_init()) \{^^J
        \ \ pr_err("error initializing usb protocol\\n");^^J
		\ \ goto error_usb;^^J
		\}^^J
		if (gb_i2c_protocol_init()) \{^^J
		\ \ pr_err("error initializing\ $\colorbox{mygray}{usb}$ protocol\\n");^^J
		\ \ goto error_i2c;^^J
		\}^^J
		\end{lstlisting} \\
        Updated		& \begin{lstlisting}[style=tblstyle]^^J
        if (gb_usb_protocol_init()) \{^^J
        \ \ pr_err("error initializing usb protocol\\n");^^J
        \ \ goto error_usb;^^J
        \}^^J
        if (gb_i2c_protocol_init()) \{^^J
        \ \ pr_err("error initializing\ $\colorbox{mygray}{i2c}$ protocol\\n");^^J
        \ \ goto error_i2c;^^J
        \}^^J
        \end{lstlisting} \\
        \bottomrule
    \end{tabularx}
\end{table}

\paragraph{LF09: Language mistakes.} 

19\% of the 900 studied log updates are caused by spelling or grammar mistakes. Example 10 (Table~\ref{tbl:rq3writing}) shows an example where the word ``\emph{synchronously}'' was misspelt as ``\emph{synchronuously}'' in the static text. Another example can be seen in commit \spverb|748ac56|. To correct a grammatical mistake, the static text was changed from ``\texttt{Failed to registered ssb SPROM handler}'' to ``\texttt{Failed to register ssb SPROM handler}''. We also noticed that there is no standardization over the use of capitalization, grammatical style,  punctuation, etc.

\paragraph{LF10: Fixing format specifiers.} 

We found 13.89\% of the 900 studied revisions are the result of using improper format specifiers in logging statements.\footnote{\url{https://www.kernel.org/doc/Documentation/printk-formats.txt}} A typical issue related to \spverb|printk| format specifier is shown in Example 11 (Table~\ref{tbl:rq3writing}). In the example, the developer decided to use the \texttt{\%pS printk} format specifier for printing symbols from direct addresses. Moreover, the explanation for this change was given as ``\emph{This is important for the ia64, ppc64 and parisc64 architectures, while on other architectures there is no difference between \%pS and \%pF. Fix it for consistency across the kernel.}'' It appears that most of this type of changes are motivated by the intention to fix build warnings, which developers should have addressed at commit-time. We found that developers often get around using proper format specifiers by resorting to unnecessary casts.\footnote{\url{https://github.com/torvalds/linux/commit/3fcb3c836ef413d3fc848288b308eb655e08d853}} This practice was rightfully reported by~\cite{chen2017antipatterns} as a logging \textit{anti-pattern}.

\paragraph{LF11: Message mistakes.} 

We identified six instances where developers made mistakes in the static messages. Similarly to LF05 (\emph{Logging wrong variables}), it seems that developers aimed at save time by copying and pasting a few snippets of code. However, some of the necessary modifications are left behind. This happens in Example 12 (Table~\ref{tbl:rq3writing}), where the developer entered an incorrect logging message in the \spverb|gpbridge_init| method, likely caused by a copy/paste mistake. It seems that the developer copied the line \spverb|if (gb_usb_protocol_init())| and corrected it. However, the debugging statements were not updated accordingly, referring to the \emph{i2c} protocol, as in the updated statement.

\paragraph{LF12: Formatting issues.}

Developers fixed formatting of log messages in 71 cases. Poorly formatted messages make it difficult to search for matching text in the log file. We can see this issue in commit \spverb|a790634|, where certain messages were appearing on separate lines resulting in a strange output. This was fixed using line continuations where necessary. Another issue that falls under this category is that developers frequently break log message strings over several lines to meet the checkpatch's 80 characters per line restriction. According to commit \spverb|4bd69e7b|, this is no longer considered a good practice, because it makes it more difficult to \spverb|grep| for strings in the source code.

\subsubsection{Disclosure of sensitive data}

\begin{table}
    \caption{Example of sensitive data.\label{tbl:rq3sensitiveData}}
    \setlength{\fboxsep}{0.1pt}
    \centering
    \begin{tabularx}{\textwidth}{lX}
    \toprule
        Example 13 & \textbf{drm/exynos: Print kernel pointers in a restricted form} $\vert$ linux\textbf{@}9cdf0ed \\ 
        \midrule
        Original & \begin{lstlisting}[style=tblstyle]^^J
        dev_dbg(dev, "< xfer\ $\colorbox{mygray}{\%p}$: tx len \%u, done \%u, rx len \%u, done \%u\\n", xfer, length, xfer->tx_done, xfer->rx_len, xfer->rx_done);^^J
        \end{lstlisting} \\
        Updated & \begin{lstlisting}[style=tblstyle]^^J
        dev_dbg(dev, "< xfer\ $\colorbox{mygray}{\%pK}$: tx len \%u, done \%u, rx len \%u, done \%u\\n", xfer, length, xfer->tx_done, xfer->rx_len, xfer->rx_done);^^J
        \end{lstlisting} \\        
        \bottomrule
    \end{tabularx}
\end{table}

\paragraph{LF13: Revealing kernel pointers.}

Insufficient information in log messages can delay the diagnosis process. However, revealing sensitive information such as cryptographic keys or kernel addresses can lead to information leaks.\footnote{\url{https://cwe.mitre.org/data/definitions/200.html}} We found 16 commits, which mentioned these cases. An example can be seen in commit \spverb|9cdf0ed| shown in Example 13 (Table~\ref{tbl:rq3sensitiveData}). It states in the commit message that ``\emph{Printing raw kernel pointers might reveal information which sometimes we try to hide (e.g. with Kernel Address Space Layout Randomization). Use the "\%pK" format so these pointers will be hidden for unprivileged users.}'' Other examples of such cases can be seen at CVE-2018-5995\footnote{\url{https://nvd.nist.gov/vuln/detail/CVE-2018-5995}} and CVE-2018-7273\footnote{\url{https://nvd.nist.gov/vuln/detail/CVE-2018-7273}}, where developers print kernel addresses into logs, which can allow an attacker to extract sensitive information. The problem of accidental data leakage through the misuse of logs has recently been examined by~\cite{zhou2020mobilogleak}. The authors showed that logs could reveal sensitive information in Android apps. Similar studies should be conducted for larger systems such as the Linux kernel to understand the extent of this serious problem. 

\subsubsection{RQ3: Summary}

\paragraph{\textbf{RQ3 - Findings.}} 

After manually analyzing 900 commits randomly selected from those that include fixes or improvements in logging code, we identified 13 types of logging fix, categorized into five groups: lack of information, issues in variable usage, inadequate logging configuration, writing issues, and disclosure of sensitive data. The most common type of logging fix consists of language mistakes, appearing in 19\% of the commits, followed by fixing inconsistencies (18\%), log levels (17\%), and format specifiers (13\%). The last three types of logging fix indicate a lack of logging standards. Another type of fix that appeared with a certain frequency is missing variables and statements (10\%), which suggests that developers should further consider all information that is needed for debugging when adding logging information.

\paragraph{\textcolor{black}{\textbf{RQ3 - Implications.}}}

\textcolor{black}{The findings show that there are many errors in the logging code, which  may have a significant impact on log analysis tasks. Based on these findings, we derived a list of guidelines and recommendations to help address each of the issues raised in this section. The proposed guidelines can be found in Section 5.4.}



\section{Discussion}\label{sec:discussion}

In this section, we discuss insights derived from our study. First, we focus on lessons learned associated with the pervasiveness of logging, \textcolor{black}{followed by comparison of our findings to previous similar studies}. Then, we present guidelines and point out directions for automated support for logging. \textcolor{black}{Next, we detail the feedback provided by a Linux expert regarding the findings of our study.} Finally, we discuss threats to the validity of our study.

\subsection{On the Pervasiveness of Logging}

Logging code represents 3.69\% of the Linux kernel, being pervasive in its source code. However, when we evaluate the pervasiveness of logging at the file and method levels, we found that the distribution of logging code is highly skewed, suggesting that certain aspects of the Linux kernel are more likely to have logging code than others. It would be interesting to assess the correlation between the log ratio of a method and its cyclomatic complexity in future studies. 

Although logging code corresponds to only less than 4\% of the Linux kernel code, this type of code is actively maintained. We found that 14\% of commits related to Linux kernel versions 4.3 to 5.3 involves modifications to logging code resulting in a total of 211,437 logging statements modifications, which represents 66.19\% of the total number of logging statements present in Linux kernel v5.3. Moreover, we identified that logging code deletion accounts for 29.23\% of the total modifications to logging code, in contrary to findings reported by~\cite{yuan2012characterizing}. Despite the majority of logging code deletion happens when a file is deleted, there is a non-negligible amount (44.66\%) of deletion of logging code occurring as afterthoughts. A possible explanation for this might be observed in commit \spverb|a8d5dad|, where a developer deleted two logging statements that report memory allocation failures. There is a rule in \spverb|checkpatch.pl| to check for possible unnecessary \emph{out of memory} message. However, it seems that developers do not care to fix this issue at commit time, because there is a large number of commits removing \emph{out of memory} error messages as afterthoughts. Logging statements are also deleted in cases in which they are redundant. For instance, in commit \spverb|c99a23e55|, a developer removed an error message present within error handling code when \spverb|i2c_mux_add_adapter| fails because \spverb|i2c_mux_add_adapter| already prints an error message when it fails.

We found that developers often face difficulties to specify the right log level in the first attempt, because there is no such a static verification of whether a log level is correct. For example, in commit \spverb|3b364c659|, a developer downgraded the logging message from \texttt{WARN} to \texttt{INFO} log level, and reason for this change was: ``\emph{On an embedded system it is quite possible for the bootloader to avoid configuring PCIe devices if they are not needed.}'' It is surprising that it took the developers 26 months to notice this problem.

The majority of the logging statements (55.66\%) are used within an \texttt{if} block. Moreover, most of them is used for logging errors after checking the return value of function calls. As the decision regarding logging is left to developers, we noticed many inconsistencies in the text of an error message, logging function used, and information included in an error message. Table~\ref{tbl:disc:thermal} shows an example of this situation. A call to \texttt{thermal\_zone\_device\_register()} returns a pointer to the newly created struct \texttt{thermal\_zone\_device}, and in case of error it returns an \texttt{ERR\_PTR}.\footnote{\url{https://github.com/torvalds/linux/blob/v5.3/drivers/thermal/thermal_core.c\#L1211}} All these three drivers perform a registration of a new thermal zone device and check the return value using \texttt{IS\_ERR()} macros and, in case of an error, log an error using \texttt{dev\_err()}. Even though all three drivers are performing the same action, there is no consistency among the error messages. This can prevent automated post-mortem analysis.


\begin{table}
    \caption{Lack of consistency in the text of the error messages} \label{tbl:disc:thermal}
    \setlength{\fboxsep}{0.1pt}
    \centering
    \begin{tabularx}{\textwidth}{X}
    \toprule
        {\begin{lstlisting}[style=tblstyle, title={\textbf{drivers/thermal/spear\_thermal.c}}, captionpos=b]^^J
        spear_thermal = thermal_zone_device_register(...);^^J
        if (IS_ERR\(spear_thermal)) \{^^J
        \ \ dev_err(&pdev->dev,\ $\colorbox{mygray}{"thermal zone device is NULL\textbackslash n"}$);^^J
        \ \ [...]^^J
        \}^^J
        \end{lstlisting}} \\ 
        \midrule
        {\begin{lstlisting}[style=tblstyle, title={\textbf{drivers/thermal/rcar\_thermal.c}}, captionpos=b]^^J
        priv->zone = thermal_zone_device_register(...);^^J
        if (IS_ERR(priv->zone)) \{^^J
        \ \ dev_err(dev,\ $\colorbox{mygray}{"can't register thermal zone\textbackslash n"}$);^^J
        \ \ [...]^^J
        \}^^J
        \end{lstlisting}} \\
        \midrule
        {\begin{lstlisting}[style=tblstyle, title={\textbf{drivers/thermal/st/st\_thermal.c}}, captionpos=b]^^J
        sensor->thermal_dev = thermal_zone_device_register(...);^^J
        if (IS_ERR(sensor->thermal_dev)) \{^^J
        \ \ dev_err(dev,\ $\colorbox{mygray}{"failed to register thermal zone device\textbackslash n"}$);^^J
        \ \ [...]^^J
        \}^^J
        \end{lstlisting}} \\
    \bottomrule
\end{tabularx}
\end{table}

We also observed inconsistencies in the information included when logging an error message. For example, there are many locations that report an error when there is a failure when calling the function \spverb|devm_request_irq|. However, plenty of the callers of this function do not include the \emph{irq} requested and the returned error code.\footnote{\url{https://github.com/torvalds/linux/blob/v5.3/drivers/usb/dwc2/gadget.c\#L4846}} A possible solution to avoid such inconsistencies would be centralizing error reporting rather than leaving the decision to the developers. One such a change made to get more consistent error reporting can be seen in commit \spverb|7723f4c|, the reason for this change is that ``\emph{A grep of the kernel shows that many drivers print an error message if they fail to get the irq they're looking for. Furthermore, those drivers all decide to print the device name, or not, and the irq they were requesting, or not, etc. Let's consolidate all these error messages into the API itself, allowing us to get rid of the error messages in each driver.}'' This centralization of error reporting helps reduce the size of the Linux kernel code. This also decreases the need for commits made as afterthoughts to add additional information or fixing typographical mistakes.

\subsection{Comparison to Existing Studies}

\textcolor{black}{Our study replicates existing studies that had as target systems C/C++ and Java projects. We, in turn, analyzed the Linux kernel, which is one of the greatest collaborative efforts in the computer industry. Based on these previous studies and their results, we show in} Table~\ref{tbl:comparison}  a comparison of the results obtained in our study and those from \cite{yuan2012characterizing} and \cite{chen2017characterizing}, which explore similar aspects related to the practice of logging.

\begin{table}
    \caption{Comparison of results from our study and previous work.\label{tbl:comparison}}
    \centering
    \begin{tabularx}{\textwidth}{LLLL}
        \toprule
        \textbf{Research Question} & \textbf{Our study} & \textbf{\cite{yuan2012characterizing}} & \textbf{\cite{chen2017characterizing}}\\ 
        \midrule
        \emph{RQ1: Pervasiveness of logging in Linux kernel}
        & On average, every 27 lines of code contained one line of logging code in the Linux kernel.
        & On average, every 30 lines of code contained one line of logging code in four open-source C/C++ projects. & On average, every 51 lines of code contained one line of logging code in 21 Java projects. 
         \\ \midrule
        
        \emph{RQ2: Logging code evolution} \ & Logging code is modified in 14\% of all analysed commits. & Logging code is modified in 18\% of all committed revisions. & Logging code is modified in 16\% of all committed revisions. \\ \cmidrule{2-4}
         & Logging deletions account for 29\% of the number of modifications made to logging code. & Logging deletions account for 2\% of the number of modifications made to logging code. & Logging deletions account for 26\% of the number of modifications made to logging code. \\ \midrule
        \emph{RQ3: Afterthought changes in logging code} & 30\% of analyzed commits are related to afterthought updates. & 33\% of updates to the log printing code are afterthought updates. & 59\% of updates to the log printing code are afterthought updates. \\ \cmidrule{2-4}
        & Fixing incorrect log levels accounts for 17\% of analyzed commits. & Fixing incorrect log levels accounts for 26\% of afterthought updates. & Fixing incorrect log levels accounts for 21\% of afterthought updates. \\ \cmidrule{2-4}
        & Almost 16\% of analyzed commits are related to issues in variable usage or missing variables and statements. & 27\% of the afterthought updates are related to variable logging. & 32\% of afterthought updates are related to variable logging. \\ \cmidrule{2-4}
        & Fixing language mistakes is the most frequent update to static text. & Fixing misleading information is the most frequent updates to the static text. & Fixing misleading information is the most frequent updates to the static text. \\
        \bottomrule
    \end{tabularx}
\end{table}

\subsection{Feedback from a Linux Expert}
\textcolor{black}{We reached out to a Linux development expert, who has also been very active in discussing the use of logging libraries in Linux, to obtain feedback on this study and discuss possible improvements. We also gave him an early version of the paper to review.  In general, the expert agreed with the fact that logging is not a practice that is  well governed in Linux, which explains the fact that logging varies significantly from one component of the system to another, depending on the development teams of the components. He added that most of the findings related the change of the logging code are caused by a lack of guidelines across all the development teams in Linux. He added that he had himself committed many updates to fix issues related to logging code to improve the quality of the logging statements. }

\textcolor{black}{Further, the Linux expert made three suggestions that do not only provide insights into the results, but  can also be used to drive future studies. The first suggestion is to  separate code written specifically for Linux from that ported to Linux such as some external drivers. This would perhaps provide a better picture of the logging landscape of Linux-specific code. Having said, our answer to RQ1 shows that the most logged components are "filesystem" and "core", which are specific to Linux. In other words,  it is not clear whether excluding the drivers would have an important effect on the results of RQ2 and RQ3. Besides drivers are an integral part of Linux, removing them may weaken the study. }
\textcolor{black}{
The second suggestion is related to the use of various tools (codespell, integrated spelling tests in checkpatch, coccinelle, etc) that have recently been made available to Linux developers. These tools are now recommended to improve the quality of code.  On one hand, this comment confirms our findings with respect to RQ3 where we concluded that many errors in the logging code could have been avoided using this type of tools. On the other hand, the point raised by the expert warrants a study on how these tools are used to check logging code and what the impact on the quality of logging statements would be.}
\textcolor{black}{
Finally,  the expert explained that in recent years, there has been an increase in the use of tracing tools such as ftrace to log the entry/exit of function, which justifies the decrease in the use of logging. This comment is inline with our finding in RQ2 where we showed that despite the increase of the number of LOCs across versions, the number of logging statements have decreased. We attributed this to the emergence of tracing and other debugging mechanisms as substitutes to logging on the premise of the studies of Corbet (2016) and Edge (2019). The Linux expert's opinion converges with this finding.} 
\textcolor{black}{
The input provided by the Linux expert encourages us to conduct a formal user study to gain more insights into the way logging is used in Linux. This study should build on the results obtained in this paper, which are an important first step to  understand  the  logging landscape in Linux in particular and in software engineering in general. 
}

\subsection{Logging Guidelines}

Based on the analysis of the afterthoughts changes in the logging statements in the Linux kernel source code, we derived the following guidelines that can help the maintenance of logging code. For each listed guideline, we also indicate the identified types of logging fix that serves as a foundation for it and a description. We also indicate opportunities for automation when applicable.

\noindent\textbf{A. Be precise, concise and consistent in logging statements.} (\emph{LF02}, \emph{LF07}, \emph{LF08}, \emph{LF10}, \emph{LF12}) In order for logs to be useful, they should be precise (not leaving room for ambiguity), concise (to not create huge amounts of logs with redundant information) and consistent (to facilitate post-mortem analysis). An example of specific recommendations is to \emph{not use hard-coded module or function names} and refer to \emph{complete variable names}. 

Developers should strive for writing concise logging statements to prevent redundant information. Automatic tools can be developed to detect redundancies. However, some redundancies may require domain knowledge. These can be detected by developers when reviewing the code. There are static analysis tools such as \emph{smatch}\footnote{\url{https://repo.or.cz/w/smatch.git}} and \emph{sparse}\footnote{\url{https://sparse.wiki.kernel.org/}}, that are useful at detecting incorrect format specifiers. Developers should consider using  these tools to detect this type of problems at commit time rather than fixing the corresponding code as afterthought. 

It is recommended to use a \emph{spell checker}, as provided by many IDEs. This is an initial step to prevent the need for corrections in language issues. Additional tools for grammar checking could be embedded in current IDEs. Guidelines for a common writing style should be developed and promoted. Tools such as \emph{kernelscan}\footnote{\url{https://github.com/ColinIanKing/kernelscan}} should be incorporated into build pipelines.

\noindent\textbf{B. Specify (in advance) and follow logging conventions.} (\emph{LF06}, \emph{LF08}) Even though there are existing descriptions of how the different log levels should be used in the Linux kernel, a suggestion is to further detail logging conventions. For instance, the error level should only be used in situations in which components may fail and compromise system operation. Moreover, conventions can be used to avoid redundant or missing information, and use similar static messages, because for the same situation, such as an error, different messages and variables are logged in distinct code locations. This can be aided by dedicated logging functions.

\noindent\textbf{C. Program for debuggability.} (\emph{LF01}, \emph{LF12}) To prevent posterior logging changes, we recommend that developers provide beforehand detailed information on where the failures are located and any other information relevant to failure that can facilitate debugging. For example, including error code in the logging statement, if available, is always a good idea. In addition, developers should avoid breaking log message lines to facilitate post-mortem analysis of the logs.

\noindent\textbf{D. Reason about possible variable values.} (\emph{LF03}, \emph{LF04}) NULL pointer dereferencing causes a runtime crash. Consequently, it is crucial to reason if a referred pointer can be NULL or a variable can be empty. A potential NULL pointer dereferencing can be avoided by adding automatic checks to logged variables, and function return values that are used as logging statements arguments. Furthermore, developers should only use the device or network-specific logging macros after checking that the devices were correctly initialized and registered. Static analysis tools can be used to check that all variables are initialized before they are used for logging.

\noindent\textbf{E. Be careful with copy-paste.} (\emph{LF05}, \emph{LF11}) Developers should avoid copying and pasting logging statements, as this is a source of mistakes. If the statements do not match the targeted subject, they may confuse and mislead developers when debugging and analyzing logged messages. In the case of a copy-paste of a logging statement, it is essencial to check if (i) a called logging function has been updated; (ii) the static message has been adequately changed; and (iii) referred variables are correct. 

\noindent\textbf{F. Consider security issues.} (\emph{LF13}) Tests should be put in place to verify that logs do not accidentally cause security and data privacy breaches. This effort should adhere to the broader task of ensuring the security of the Linux kernel.

\noindent\textbf{G. Review logging code.} (\emph{LF02}, \emph{LF09}) To avoid many of the raised issues, such as imprecise, redundant, and inconsistent logging statements as well as copy-paste mistakes, can be detected at commit time during a code review process. Therefore, we recommend logging code to be carefully reviewed. This task can be done before accepting the commits. 

\noindent\textbf{H. Use recommender systems.} (\emph{LF01}, \emph{LF05}, \emph{LF06}, \emph{LF07})
There is an ongoing research in developing recommendation tools  to assist developers in logging. For example, \citet{panthaplackel2020learning} proposed a deep-learning-based approach to automatically update the static text in the logging statements based on changes made to the surrounding source code. \citet{fu2014developers} conducted a study at Microsoft where  they showed that it is possible to use machine learning techniques to predict with a good accuracy where to log. Developers should consider including these tools, if shown effective in the context of the Linux kernel, to predict log levels,  variables that need to be logged, static analysis of logging statements, etc.  

\subsection{Implications for Future Research}

Our study allowed us to understand how logging is present in the Linux kernel and to identify issues that caused the evolution of logging statements. The analysis of collected data revealed facts related to logging in the Linux kernel, leaving room for future research work. We discuss directions for future studies as follows. 

The analysis of the pervasiveness of the logging in the Linux kernel showed that the log ratio in the many Linux subsystems and components varies. This can be explained by various arguments, such as: (i) long-lived less-changed components may be stable and are the cause of few bugs, so they do not need to be as logged as other components; (ii) logging can have an impact on system performance and, therefore, certain code parts may be less logged to prevent performance decay; and (iii) different parts of the code may be maintained by different teams of developers, with diverging logging practices. Based on the data collected for this study, it is not possible to reach a conclusion regarding the rationale for the current logging decisions and their variability across the different Linux subsystems and files. This calls for future studies that investigate this. Possible types of studies that are suitable for testing these hypotheses are:

\begin{itemize}
    \item a user study in which developers think out loud, justifying the logging statements that they add, modify or delete;
    \item interviews with developers immediately after they commit a code change that involves logging changes; or
    \item a qualitative study of code review comments, in which there are requests for changes in logging statements with an accompanying rationale for the change.
\end{itemize}

With respect to our logging guidelines, a future study can be conducted to evaluate their impact on the evolution of logging practices. The presented guidelines were derived using the principles of grounded theory---the guidelines are grounded on the conducted qualitative analysis. Although there are supporting evidences for each derived guideline, it is important to assess if their adoption in the Linux kernel (or a software project) can have a positive impact on the standardization of the logging statements, reducing the need for constant modifications due to inadequate logging. An experimental study in which the intervention is the use of our derived guidelines is suitable for answering this question.

\subsection{Threats to Validity}\label{sec:threats}

\paragraph{Internal Validity.}

Threats to internal validity are associated with factors that may impact our results. In this study, a source of bias is the automated data collection process. To identify logging statements, we rely on the semantic patterns specified by~\cite{tschudin20153l}. However, this approach may not identify statements that lack variability in their use. To mitigate this issue, we manually examined macros containing calls to basic functions, such as \texttt{printk()}, \texttt{pr\_*()}, and \texttt{dev\_*()}, in order to identify missed logging functions. The set of functions collected by combining both processes was thus manually reviewed with the aim of eliminating obvious false positives. In addition, although we believe that the list of logging functions extracted with this approach is a very comprehensive, we cannot guarantee 100\% coverage. To mitigate this threat, we selected randomly 20 files and checked manually their content to see if we  missed any logging functions. We found that our list  covered all the logging functions invoked in these files. We also want to emphasize that we do not think that the missing of some logging functions affects much the conclusions of this study.  For RQ1 (the pervasiveness of logging), we already established that logging is pervasive in Linux (on average for every 27 lines of code, there is one logging statement), so finding more logging statements can only confirm this fact. For RQ2-The evolution of logging code, we found  that logging code is constantly maintained by Linux developers (14\% of commits modify logging statements). Additional logging statements will eventually confirm this finding. The nature of changes are due to the need to improve the quality of the logging output by either enhancing precision,conciseness, or consistency.  RQ3 (analysis of afterthought log changes) is based on a sample of commits and is not affected by the number of  logging statements.

In order to automatically identify and classify changes made to logging code, we use a pipeline that consists of GumTree~\citep{FalleriMBMM14} and a script specifically developed for this study. To mitigate possible misclassifications, we manually inspected a sample of 100 randomly selected modifications, which showed an accuracy of 98\% on classified modifications. Although 2\% have been misclassified, given the amount of analyzed code, it can be considered noise in the data.

Another threat to this study is the manual classification of log-related commits. We manually examined all commits by their title and message whenever necessary. However, we cannot eliminate the possibility that errors may have occurred during this manual analysis. We thus do not claim that our dataset is complete. This threat is mitigated by the fact that our goal was to collectively study the nature of the problematic logging code rather than collecting every possible revision made to fix or improve logging. Because we perform a qualitative analysis in order to categorize changes made to logging code and the nature of problematic logging code, researcher bias also becomes a threat to the internal validity of our study. In this kind of analysis, results are associated with researcher interpretation of the data. Therefore, to mitigate it, every classification that raised questions were discussed by the authors until they reached an agreement. In addition, all the results were largely discussed and validated among all authors.

\paragraph{External Validity.}

Threats to external validity are related to which extent our results can be generalized. One threat to external validity is due to the fact that we only examined 22 releases (v4.3 - v5.3) of the Linux kernel  to answer part of RQ2 where we  measure the evolution of logging code in these releases. A larger dataset may provide different results. This said, we still think our dataset is representative for the study as a whole since our objective is not to uncover all the problems related to logging in Linux but rather to provide insight on the practice of logging in software development by looking at how Linux developers, even in a narrower scope, use logging. In addition, the fact that we focused solely on the Linux kernel project, which is known for having its own development culture, may be a threat to external validity. We believe that this threats is mitigate by the fact that Linux is a large scale, open-source project maintained by developers from different companies, which constitutes a representative sample of C/C++ projects.


\section{Conclusion}\label{sec:conclusion}

In this paper, we presented an empirical study that aims to shed light into existing logging practices in the Linux kernel. Although the logging code accounts for 3.73\% of the total source code in the Linux kernel, we observed that the distribution of logging code is skewed when evaluated at file and program construct level. Future studies may employ survey methods to understand the circumstances in which logging is needed and the rationales of logging activities. We also found that 32.37\% of the log-level changes are between error and debug log levels, suggesting that it is also difficult for Linux developers to decide between fatal errors and errors that are recoverable. We observed that the major causes underlying fixes to problematic logging code occurring in the Linux kernel are language mistakes, appearing in 19\% of the commits. Other common causes for changes are fixing inconsistencies (18\%), log levels (17\%), and format specifiers (13\%). Based on our results, we discussed insights associated with the pervasiveness of logging code in the Linux kernel. In addition, we derived eight practical guidelines that can help developers to maintain logging code, including the need for automated support for logging.

As future work, we aim to perform an in-depth qualitative analysis of logging code to further understand common patterns in logging statements. Our present study targeting the Linux kernel, complemented by this future qualitative analysis, will be the basis for the next generation of tools to provide automated support to develop and evolve logging code.

\section*{Acknowledgements}
Abdelwahab Hamou-Lhadj and Keyur Patel would like to thank Ericsson Global Artificial Intelligence Accelerator (GAIA) Group in Montreal and MITACS for supporting this project (Grant Number: IT15986). 
Ingrid Nunes would like to for CNPq grants ref. 313357/2018-8 and ref. 428157/2018-1, and the Coordena\c{c}\~{a}o de Aperfei\c{c}oamento de Pessoal de N\'{i}vel Superior - Brasil (CAPES) - Finance Code 001. João Faccin would like to acknowledge the support of the National Council for Scientific and Technological Development of Brazil (CNPq) (grant ref. 141840/2016-1), and the support of the Government of Canada through the 
Emerging Leaders in the Americas Program (ELAP) program.

\bibliographystyle{spbasic}
\bibliography{reference}

\end{document}